\documentclass[twocolumn,aps,letterpaper,showpacs]{revtex4}
\usepackage[T1]{fontenc}
\usepackage{color}
\usepackage{graphicx}
\usepackage[unicode=true, pdfusetitle,
 bookmarks=true,bookmarksnumbered=false,bookmarksopen=false,
 breaklinks=false,pdfborder={0 0 0},backref=false,colorlinks=true]
 {hyperref}
\hypersetup{
 linkcolor=blue,citecolor=green}
 
\makeatletter

\providecommand{\tabularnewline}{\\}

\@ifundefined{textcolor}{}
{%
 \definecolor{BLACK}{gray}{0}
 \definecolor{WHITE}{gray}{1}
 \definecolor{RED}{rgb}{1,0,0}
 \definecolor{GREEN}{rgb}{0,1,0}
 \definecolor{BLUE}{rgb}{0,0,1}
 \definecolor{CYAN}{cmyk}{1,0,0,0}
 \definecolor{MAGENTA}{cmyk}{0,1,0,0}
 \definecolor{YELLOW}{cmyk}{0,0,1,0}
 }

\usepackage[OT1]{fontenc}
\usepackage{aas_macros}
\definecolor{green}{rgb}{0,.7,0}
\definecolor{blue}{rgb}{.15,.15,1}

\makeatother

\begin{document}

\title{Stationary, Axisymmetric Neutron Stars\\
with Meridional Circulation in General Relativity}

\author{Reiner Birkl$^1$, Nikolaos Stergioulas$^2$, Ewald M\"{u}ller$^1$}

\affiliation{$^1$Max-Planck-Institut f\"{u}r Astrophysik, 85741 Garching, Germany\\
$^2$Aristotle University of Thessaloniki, 54124 Thessaloniki,
Greece}
\begin{abstract}
We present the first stationary, axisymmetric neutron star models
with meridional circulation in general relativity. For that purpose,
we developed GRNS, a new code based on a fixed point iteration. We
find a two-dimensional set of meridional circulation modes, which
differ by the number of vortices in the stream lines of the neutron
star fluid. For expected maximal meridional circulation velocities
of about $1000\,{\rm km}/{\rm s}$, the vortices cause surface deformations
of about a percent. The deformations depend on the shape of the vortices
close to the surface and increase with the meridional circulation
velocity. We also computed models of rotating neutron stars with meridional
circulation, where neither the surface rotates nor does the rotation
velocity exceed the circulation velocity.
\end{abstract}

\pacs{97.60.Jd, 04.20.-q, 95.30.Lz}

\maketitle

\section{Introduction}

The shape of a neutron star depends on the motion of the stellar fluid.
This motion can be decomposed into two components with respect to
the rotation axis, which contains the center of mass and points in
the direction of the total angular momentum. The first component is
differential rotation around the rotation axis. This kind of motion,
which leads to a flattening of the neutron star surface, has been
investigated intensively, also in general relativity \cite{1995.Nick,1998.Nick}.
The second component is meridional circulation, where the flow occurs
in meridional planes, i.e. planes containing the rotation axis. Its
influence, which may play a role for new-born neutron stars, is not
yet well understood. Earlier investigations involve severe limitations,
as they were either restricted to the Newtonian framework \cite{1986.Ewald}
or they were of perturbative nature only \cite{2004.Ioka}. We are
interested in how meridional circulations of arbitrary strength may
influence the shape of a general relativistic neutron star. Moreover,
contrary to the earlier studies considering non-perturbative fluid
motions, where in each case only one of the two kinds of motions was
considered alone, we also investigate the combined effect of differential
rotation and meridional circulation.

For that purpose, we developed a new program, called GRNS (=Generally
Rotating Neutron Star), as the complexity of Einstein's field equation
makes a numerical approach to our investigation unavoidable. GRNS
combines the methods applied in two programs created by two of us
more than a decade ago. The first one is RNS \cite{1995.Nick,1998.Nick}
, a general relativistic code for computing models of rapidly rotating
neutron stars. This code is based on a fixed point iteration, and
the theoretical considerations of Komatsu \textit{et al.~}\cite{1989.Komatsu}.
However, RNS is limited to differentially rotating configurations
and does not allow for meridional circulation. The situation is just
opposite for the second program (E. M\"{u}ller \cite{1986.Ewald}),
which computes meridional circulation in non-rotating Newtonian configurations.
It uses a Newton-Raphson iteration scheme to solve for the stationary
equilibrium configurations, and describes the flow by means of a scalar
function, the so-called stream function. Below we will show how the
stream function method can be extended to general relativity. We do
not use the approach of Komatsu \textit{et al.~}\cite{1989.Komatsu}
to describe the effects of meridional circulation on the curvature
of spacetime, but rely on the more general, theoretical investigation
of Gourgoulhon \textit{et al.~}\cite{1993.Gourgoulhon}, who formulated
Einstein's field equation as a set of covariant Poisson equations.
The latter can be rewritten as Poisson equations in flat space such
that Green functions can be used to solve them. We have incorporated
these theoretical extensions into GRNS, including the fixed point
iteration of RNS, and a modified Tolman-Oppenheimer-Volkoff solution
as the initial guess. That way, GRNS is able to handle differentially
rotating neutron stars with meridional circulation.

GRNS uses a simplified neutron star model. Considering the neutron
star during a sufficiently short time interval after its formation,
it is an appropriate approximation to assume stationarity. In addition,
we limit ourselves to axisymmetric configurations. For simplicity,
the stellar matter in our simulations is assumed to be a perfect fluid
described by a barotropic equation of state, although the theoretical
formulation presented in this paper is kept general enough to deal
with arbitrary equations of state. We further assume that the neutron
star has no substructure (i.e., crust, core, etc.), and that its chemical
composition is homogeneous. Finally, we neglect the influence of magnetic
fields.

The paper is organized as follows. In Sec. \ref{sec:Theory}, we rewrite
the covariant Poisson equations of Gourgoulhon \textit{et al.~}\cite{1993.Gourgoulhon}
to flat-space Poisson equations and extend the Newtonian stream function
method \cite{1986.Ewald} to general relativity. Intermediate steps
during these computations are listed in the Appendix. Sec. \ref{sec:Numerics}
gives a closer look at the GRNS code explaining the fixed point iteration
scheme and showing how the flat-space Poisson equations can be solved
by means of Green functions. Actual neutron star models are presented
in Sec. \ref{sec:Results}, and we conclude our paper in Sec. \ref{sec:Conclusions}.

\section{Theory\label{sec:Theory}}

\subsection{Notations and conventions}

We use geometrized units for the theoretical computations, where the
speed of light $c$ and the gravitational constant $G$ are set equal
to unity. The results obtained from neutron star simulations are presented
in the more appropriate cgs units. The general relativistic formalism
adopts Einstein's sum convention and is kept as close as possible
to the conventions chosen in Gourgoulhon \textit{et al.~}\cite{1993.Gourgoulhon}.
However, we make a slightly different choice for the range of tensor
indices. In our specific choice of coordinates, concrete indices for
4-tensors, 3-tensors and 2-tensors are used as follows: \begin{eqnarray}
\alpha,\beta,... & \in & \{t,r,\theta,\phi\},\nonumber \\
a,b,... & \in & \{r,\theta,\phi\},\nonumber \\
m,n,... & \in & \{r,\theta\},\label{eq:1}\end{eqnarray}
with the time $t$, the radius $r$, and the angles $\theta$ and
$\phi$. Following Misner \textit{et al.~}\cite{1973.Wheeler}, we
choose $\left(-,+,+,+\right)$ as the signature of the metric.

\subsection{Basic fields\label{sub:Basic-fields}}

We begin our theoretical considerations by defining a set of physical
fields, the so-called basic fields, which determine the state of the
neutron star uniquely. Due to the assumption of stationarity and axisymmetry,
these and all other fields depend on only two coordinates, the radius
$r$ and the polar angle $\theta$. There are two types of basic fields.
The first set describes the curvature of spacetime, and its fields
are denoted basic geometry fields. The second set are hydrodynamic
fields, called basic matter fields, which specify the properties of
the stellar fluid. In the following, we will define the basic geometry
fields without giving an explanation of how these fields come into
being. For that purpose, we refer to Gourgoulhon \textit{et al}.~\cite{1993.Gourgoulhon}.

The curvature of spacetime is described by the metric $g_{\alpha\beta}$.
Using a $\left(2+1\right)+1$ decomposition of spacetime, the ten
independent components of the $4$-metric $g_{\alpha\beta}$ are rewritten
to new fields in Gourgoulhon \textit{et al.~}\cite{1993.Gourgoulhon}.
These fields consist of two scalars fields, the $3$-lapse $N$ and
the $2$-lapse $M$, two vector fields, the $3$-shift $N^{a}$ and
the $2$-shift $M^{m}$, and the $2$-metric $k_{mn}$, respectively:\[
\left(\begin{array}{cc}
g_{tt} & g_{tb}\\
g_{at} & g_{ab}\end{array}\right)=\left(\begin{array}{cc}
N_{c}N^{c}-N^{2} & -N_{b}\\
-N_{a} & h_{ab}\end{array}\right),\]
with the $3$-metric\[
\left(\begin{array}{cc}
h_{mn} & h_{m\phi}\\
h_{\phi n} & h_{\phi\phi}\end{array}\right)=\left(\begin{array}{cc}
k_{mn} & -M_{m}\\
-M_{n} & M^{2}+M_{p}M^{p}\end{array}\right)\]
and\begin{eqnarray*}
N_{a} & = & h_{ab}N^{b}\\
M_{m} & = & k_{mn}M^{n}.\end{eqnarray*}
Using a special choice of coordinates, which are denoted as MTCMA
(maximal time slicing - conformally minimal azimuthal slicing) in
Gourgoulhon \textit{et al.~}\cite{1993.Gourgoulhon}, these authors
write the $2$-metric $k_{mn}$ such that the only non-vanishing components
are $k_{rr}=A^{2}$ and $k_{\theta\theta}=A^{2}r^{2}$, where $A$
shall be called the $2$-conformal factor. Adopting the definitions
\cite{1993.Gourgoulhon}\begin{eqnarray}
\nu & = & \ln N,\label{eq:14}\\
\alpha & = & \ln A,\label{eq:16}\\
\beta & = & \ln\frac{M}{r\sin\theta},\label{eq:15}\end{eqnarray}
we then call the eight functions\begin{equation}
\nu,\alpha,\beta,M^{r},M^{\theta},N^{r},N^{\theta},N^{\phi}\label{eq:2}\end{equation}
basic geometry fields. In contrast, for circular spacetimes (i.e.
without meridional circulation), only four geometry fields are required.

The neutron star fluid is described by a stress-energy tensor\begin{equation}
T_{\alpha\beta}=\mbox{\ensuremath{\left(\epsilon+p\right)}}u_{\alpha}u_{\beta}+pg_{\alpha\beta},\label{eq:84}\end{equation}
with the total energy density $\epsilon$, pressure $p,$ and $4$-velocity
$u^{\alpha}$. Due to the constraint\begin{equation}
g_{\alpha\beta}u^{\alpha}u^{\beta}=-1,\label{eq:22}\end{equation}
the $4$-velocity $u^{\alpha}$ possesses only three independent degrees
of freedom. One of them is the specific angular momentum, i.e. the
$\phi$-component of\begin{equation}
l_{a}=-\frac{u_{a}}{u_{t}}.\label{eq:21}\end{equation}
Under our symmetry assumptions (Sect. \ref{sub:Stream-function}),
the other two degrees of freedom can be expressed in terms of a stream
function $\psi$, an approach known to be working fine in Newtonian
gravity (equations (2) and (3) in Eriguchi \textit{et al.~}\cite{1986.Ewald}).
Further below (Sect. \ref{sub:Stream-function}), we will demonstrate
that this approach holds in general relativity, too. There, we will
also explain how the $4$-velocity $u^{\alpha}$ can be obtained from
the stream function $\psi$ and the specific angular momentum $l_{\phi}$.
However, by convention we do not call the stream function $\psi$
a basic field, but the modified stream function~\cite{1986.Ewald}\begin{equation}
\chi_{0}=\frac{\psi}{r\sin\theta}.\label{eq:88}\end{equation}
So, we refer to\begin{equation}
\epsilon,p,\chi_{0},l_{\phi}\label{eq:3}\end{equation}
as basic matter fields. Actually, we have chosen the twelve basic
fields (\ref{eq:2}) and (\ref{eq:3}) in such a manner that they
vanish for an empty Minkowski spacetime, which is advantageous for
our numerical method.

\subsection{Geometry equations\label{sub:Geometry-equations}}

Each of the basic fields (\ref{eq:2}) and (\ref{eq:3}) is specified
by a field equation. In this section, we will address the basic geometry
fields, which are governed by Einstein's field equation\begin{equation}
G_{\alpha\beta}=8\pi T_{\alpha\beta},\label{eq:4}\end{equation}
where $G_{\alpha\beta}$ is the Einstein tensor. Unfortunately, this
equation is written in such a compact manner that it is practically
impossible to directly solve for the basic geometry fields (\ref{eq:2}).
Therefore, we use the theoretical consideration of Gourgoulhon \textit{et
al.~}\cite{1993.Gourgoulhon}, who rewrite equation (\ref{eq:4})
as a set of eight covariant Poisson equations, each of the form\begin{equation}
\Delta_{{\rm cov}}\Phi=S,\label{eq:5}\end{equation}
where $\Delta_{{\rm cov}}$, $\Phi$, and $S$ are a covariant Laplacian,
a potential, and a source, respectively. In that paper, the covariant
Poisson equations are labelled (B3-B7). Trying to validate these equations,
we found that the three equations (B3), (B4a) and (B4b) contain minor
errors. In Appendix \ref{sec:Corrected-geometry-equations}, we show
the corrected versions of these equations.

Let us now have a closer look at the covariant Laplacian $\Delta_{{\rm cov}}$
appearing in equation (\ref{eq:5}). This operator is the scalar product
of two covariant derivatives. Such derivatives can be split into a
partial derivative and corrections arising from the curvature of spacetime,
expressed in terms of the connection coefficients. In a similar manner,
it is possible to express a covariant Laplacian $\Delta_{{\rm cov}}$
in terms of the corresponding flat-space Laplacian $\Delta_{{\rm flat}}$
and curvature corrections. That way, equation (\ref{eq:5}) can be
rewritten as\[
\Delta_{{\rm flat}}\Phi=S',\]
with some new source $S'$. In this way, we arrive at the following
flat-space versions of equations (B3-B7) in \cite{1993.Gourgoulhon}:\begin{eqnarray}
^{3}\Delta\nu & = & S_{\nu},\label{eq:47}\\
^{3}\Delta_{\;\; b}^{a}N^{b} & = & S_{N}^{a},\label{eq:48}\\
^{2}\Delta\left[r\sin\theta\left(\beta+\nu\right)\right] & = & S_{\beta},\label{eq:6}\\
^{2}\Delta_{\;\;\; n}^{m}\left[e^{2\left(\alpha+\nu\right)}M^{n}\right] & = & S_{M}^{m},\label{eq:7}\\
^{2}\Delta\left(\alpha+\nu\right) & = & S_{\alpha}.\label{eq:8}\end{eqnarray}
The flat-space Laplacians $^{2}\Delta$, $^{3}\Delta$, $^{2}\Delta_{\;\;\; n}^{m}$
and $^{3}\Delta_{\;\; b}^{a}$ are listed in Appendix \ref{sec:Flat-space-Laplacians},
and the rather lengthy sources $S_{\nu}$, $S_{\alpha}$, $S_{\beta}$,
$S_{M}^{m}$ and $S_{N}^{a}$ in Appendix \ref{sec:Sources}, respectively.
Modifying the sources $S_{\beta}$, $S_{M}^{m}$, and $S_{\alpha}$
properly, it is possible to rewrite equations (\ref{eq:6}-\ref{eq:8})
such that the Laplacians in these equations act directly on the basic
geometry fields $\beta$, $M^{m}$, and $\alpha,$ but not on expressions
containing these fields. However, the numerical stability of the fixed
point iteration is highly sensitive to such changes, and in that case
the iteration would diverge. By experimenting with the GRNS code,
we found that the choice made in equations (\ref{eq:6}-\ref{eq:8})
produces convergent results.

For rotating neutron stars, we achieved convergence only by setting
the source $S_{M}^{m}$ equal to zero on the grid lines adjacent to
the rotation axis. Even when increasing the grid resolution, only
the adjacent grid lines have to be modified, and thus setting the
source equal to zero is a valid procedure.

\subsection{Matter equations}

The basic matter fields (\ref{eq:3}) are fixed by the equation of
state and the equation of general relativistic hydrodynamics\begin{equation}
\nabla_{\beta}T^{\alpha\beta}=0,\label{eq:9}\end{equation}
where $\nabla_{\alpha}$ is the covariant derivative. As will become
apparent later (Sect. \ref{sub:Poisson-equation}), for a stationary
configuration, the equations of hydrodynamics are integrable under
the assumption of a barotropic fluid, where the total energy density
$\epsilon$ is a function $\epsilon\left(p\right)$ of the pressure
$p$ only.

Similar to Einstein's field equation (\ref{eq:4}), equation (\ref{eq:9})
is again too compact to be solved directly. To overcome this problem,
we introduce the projector\[
q_{\beta}^{\alpha}=\delta_{\beta}^{\alpha}+u^{\alpha}u_{\beta}\]
orthogonal to the $4$-velocity $u^{\alpha}$, where $\delta_{\beta}^{\alpha}$
is the Kronecker symbol. Then, we project equation (\ref{eq:9}) in
the direction parallel and orthogonal to the $4$-velocity:\begin{eqnarray*}
u_{\alpha}\nabla_{\beta}T^{\alpha\beta} & = & 0,\\
q_{\alpha}^{\gamma}\nabla_{\beta}T^{\alpha\beta} & = & 0,\end{eqnarray*}
and bring these two equations in the usual form\begin{eqnarray}
\nabla_{\alpha}\left[\left(\epsilon+p\right)u^{\alpha}\right] & = & u^{\alpha}\nabla_{\alpha}p,\label{eq:10}\\
\ensuremath{\left(\epsilon+p\right)u^{\beta}\nabla_{\beta}u^{\alpha}} & = & -q^{\alpha\beta}\nabla_{\beta}p,\label{eq:11}\end{eqnarray}
which are the energy equation and the general relativistic Euler equation.
In the subsequent sections we will extend the Newtonian stream function
method \cite{1986.Ewald} to general relativity by a further reformulation
of equations (\ref{eq:10}) and (\ref{eq:11}).

\subsection{Stream function\label{sub:Stream-function}}

The stream function $\psi$ is introduced to solve the energy equation
(\ref{eq:10}). In Appendix \ref{sec:Energy-equation}, we show that
due to our symmetry assumptions equation (\ref{eq:10}) can be written
as the vanishing flat-space $3$-divergence\begin{equation}
\partial_{m}j^{m}+\frac{2}{r}j^{r}+\cot\theta j^{\theta}=0\label{eq:19}\end{equation}
of the $3$-vector\begin{equation}
j^{a}=\varrho u^{a},\label{eq:25}\end{equation}
with \cite{1989.Komatsu}\begin{eqnarray}
\varrho & = & A^{2}e^{\gamma}\left(\epsilon+p\right)u_{t},\nonumber \\
\gamma & = & \beta+\nu\label{eq:24}\end{eqnarray}
(note that the symbol $\varrho$ is different than the rest mass density
$\rho$ introduced later). Then, similar to the Newtonian stream function
method~\cite{1986.Ewald}, equation (\ref{eq:19}) is automatically
satisfied by the relation\begin{equation}
\left(\begin{array}{c}
u^{r}\\
u^{\theta}\end{array}\right)=\frac{1}{\varrho r^{2}\sin\theta}\left(\begin{array}{c}
\partial_{\theta}\psi\\
-\partial_{r}\psi\end{array}\right).\label{eq:20}\end{equation}
This relation reduces two degrees of freedom of the $4$-velocity
$u^{\alpha}$ to the one of the stream function $\psi$ (which has
the units $[g/s]$). In the Newtonian limit, where $u_{t}=-1$, these
two degrees of freedom correspond to the two components $u^{m}$.
Unfortunately, in general relativity, the situation is somewhat more
complicated, because the components $u^{m}$ together with the third
degree of freedom $u^{\phi}$ also appear on the right hand side of
equation (\ref{eq:20}), hidden in the quantity $u_{t}$. However,
using equation (\ref{eq:20}), the constraint (\ref{eq:22}), and
the definition (\ref{eq:21}) it is straightforward to show that the
$4$-velocity $u^{\alpha}$ can still be computed from the stream
function $\psi$ and the specific angular momentum $l_{\phi}$.

\subsection{Specific angular momentum}

Taking stationarity and axisymmetry into account, a short calculation
shows that the general relativistic Euler equation (\ref{eq:11})
can be written as\begin{equation}
\frac{\partial_{\alpha}p+u_{\alpha}u^{m}\partial_{m}p}{\epsilon+p}=\frac{1}{2}u^{\beta}u^{\gamma}\partial_{\alpha}g_{\beta\gamma}-u^{m}\partial_{m}u_{\alpha}.\label{eq:23}\end{equation}
We recall that this equation is a projection orthogonal to the $4$-velocity
$u^{\alpha}$ and has thus only three independent degrees of freedom.
In the following, we will therefore consider only the three spatial
components of equation (\ref{eq:23}) by setting $\alpha=a$ in that
equation. However, we use the temporal component $\alpha=t$, shown
in equation (\ref{eq:26}), to rewrite these spatial components as\begin{equation}
\frac{\partial_{a}p}{\epsilon+p}=\frac{1}{2}u^{\beta}u^{\gamma}\partial_{a}g_{\beta\gamma}+u_{t}u^{m}\partial_{m}l_{a}.\label{eq:27}\end{equation}

Let us start with the azimuthal component of equation (\ref{eq:27}).
For that purpose, we set $a=\phi$ and being aware that all azimuthal
derivatives $\partial_{\phi}...$ vanish due to axisymmetry, we obtain\begin{equation}
u^{m}\partial_{m}l_{\phi}=0,\label{eq:28}\end{equation}
which extends equation (12) of Eriguchi \textit{et al.~}\cite{1986.Ewald}
to general relativity. To solve equation (\ref{eq:28}), we discern
the three cases\begin{eqnarray*}
(A) & u^{m}=0 & \mbox{everywhere,}\\
(B) & u^{m}\neq0 & \mbox{somewhere},\\
(C) & u^{m}\neq0 & \mbox{everywhere}.\end{eqnarray*}
In case (A), for which equation (\ref{eq:28}) is satisfied automatically,
the neutron star can be differentially rotating, but without any meridional
circulation. We are not interested in this case, because it has already
been investigated with the RNS code \cite{2004.Nick,1998.Nick,1995.Nick}.
Case (B) allows a meridional circulation, but not everywhere in the
neutron star. In this paper, we do not investigate such solutions
nor do we analyze whether they exist at all. Instead, we focus on
case (C), where meridional circulation is present everywhere in the
star. For that case, we use equation (\ref{eq:20}) and rewrite equation
(\ref{eq:28}) as\[
\partial_{\theta}\psi\partial_{r}l_{\phi}-\partial_{r}\psi\partial_{\theta}l_{\phi}=0.\]
This equation is satisfied by\begin{equation}
l_{\phi}=L\left(\psi\right),\label{eq:32}\end{equation}
where $L\left(\psi\right)$ is an arbitrary function of the stream
function $\psi$ \cite{1974.Roxburgh}.

\subsection{Poisson equation for modified stream function\label{sub:Poisson-equation}}

We proceed with the meridional components of equation (\ref{eq:27}),
which can be written as\begin{equation}
\frac{\partial_{m}p}{\epsilon+p}+\frac{1}{2}\partial_{m}\ln\left(u_{t}\right)^{2}=rwu_{t}\left(\begin{array}{c}
u^{\theta}\\
-u^{r}\end{array}\right)-u_{t}u^{\phi}\partial_{m}l_{\phi}\label{eq:30}\end{equation}
according to Appendix \ref{sec:Meridional-components}, where the
quantity $w$ is defined in equation (\ref{eq:33}). This expression
extends equations (7) and (8) of Eriguchi \textit{et al.~}\cite{1986.Ewald}
to general relativity. Our limitation to a barotropic equation of
state allows us to write the left hand side of equation (\ref{eq:30})
in the form of a gradient. To this end, we introduce the heat function
\cite{1993.Gourgoulhon2}\begin{equation}
H\left(p\right)=\int_{0}^{p}\frac{{\rm d}p'}{\epsilon\left(p'\right)+p'}.\label{eq:45}\end{equation}
For the right hand side, we use equations (\ref{eq:20}) and (\ref{eq:32}).
Then, equation (\ref{eq:30}) reads\begin{eqnarray}
 &  & \partial_{m}\left[H\left(p\right)+\frac{1}{2}\ln\left(u_{t}\right)^{2}\right]\nonumber \\
 & = & -u_{t}\left(\frac{w}{\varrho r\sin\theta}+u^{\phi}L'\left(\psi\right)\right)\partial_{m}\psi,\label{eq:44}\end{eqnarray}
which implies that the right hand side must be a gradient, too, for
obtaining a first integral of motion. This leads to the condition\begin{equation}
w=\varrho r\sin\theta\left(\frac{f\left(\psi\right)}{u_{t}}-u^{\phi}L'\left(\psi\right)\right),\label{eq:34}\end{equation}
with an arbitrary function $f\left(\psi\right)$. Examining equations
(\ref{eq:33}), (\ref{eq:21}), and eventually (\ref{eq:20}), we
realize that the quantity $w$ contains second derivatives of the
stream function $\psi$. Actually, as shown in the Appendix \ref{sec:Asymmetric-Laplacian},
it turns out that there is a Laplacian hidden in that quantity, and
that condition (\ref{eq:34}) is equivalent to\begin{equation}
\frac{1}{\cos\phi}{}^{3}\Delta\left(\chi_{0}\cos\phi\right)=S_{\chi_{0}},\label{eq:35}\end{equation}
where the modified stream function $\chi_{0}$ (\ref{eq:88}) is used,
and the source $S_{\chi_{0}}$ is given in Appendix \ref{sec:Sources}.
This flat-space Poisson equation differs from that presented in Sect.
\ref{sub:Geometry-equations} by the appearance of the angle $\phi$.

\subsection{Pressure}

We now return to equation (\ref{eq:44}) and insert condition (\ref{eq:34}).
Then, by integration we obtain\begin{equation}
\frac{1}{2}\ln\left(u_{t}\right)^{2}+H\left(p\right)+I(\psi)=C,\label{eq:46}\end{equation}
with the ancillary function\[
I\left(\psi\right)=\int_{0}^{\psi}{\rm d}\psi'f\left(\psi'\right)\]
and the integration constant $C$. The heat function $H\left(p\right)$
appearing in equation (\ref{eq:46}) is invertible, because the total
energy density $\epsilon$ and the pressure $p$ are both positive
quantities, and thus also the integrand of equation (\ref{eq:45}),
such that the heat function $H\left(p\right)$ is strictly monotonous.
Denoting the inverse heat function as $H^{-1}\left(p\right)$, equation
(\ref{eq:46}) the pressure can be expressed as\begin{equation}
p=H^{-1}\left(H\left(p_{{\rm c}}\right)+\frac{1}{2}\ln\left(\frac{u_{t}^{{\rm c}}}{u_{t}}\right)^{2}+I(\psi_{{\rm c}})-I(\psi)\right),\label{eq:49}\end{equation}
with the central pressure $p_{{\rm c}}$, the central stream function
$\psi_{{\rm c}}$, and the central covariant temporal component $u_{t}^{{\rm c}}$
of the 4-velocity, respectively. In general, the above inversion is
performed numerically.

\section{Numerics\label{sec:Numerics}}

\subsection{Fixed point iteration\label{sub:Fixed-point-iteration}}

\begin{figure}
\noindent \centering{}\hspace{-.3cm}%
\begin{minipage}[c]{0.1\columnwidth}%
\noindent \begin{center}
\includegraphics[scale=0.8]{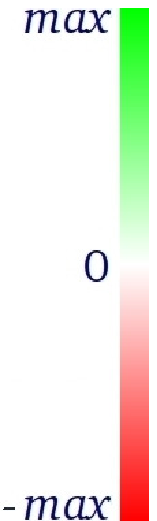}\\
\vspace{.7cm}
\par\end{center}%
\end{minipage}\put(33,0){\includegraphics[scale=0.35]{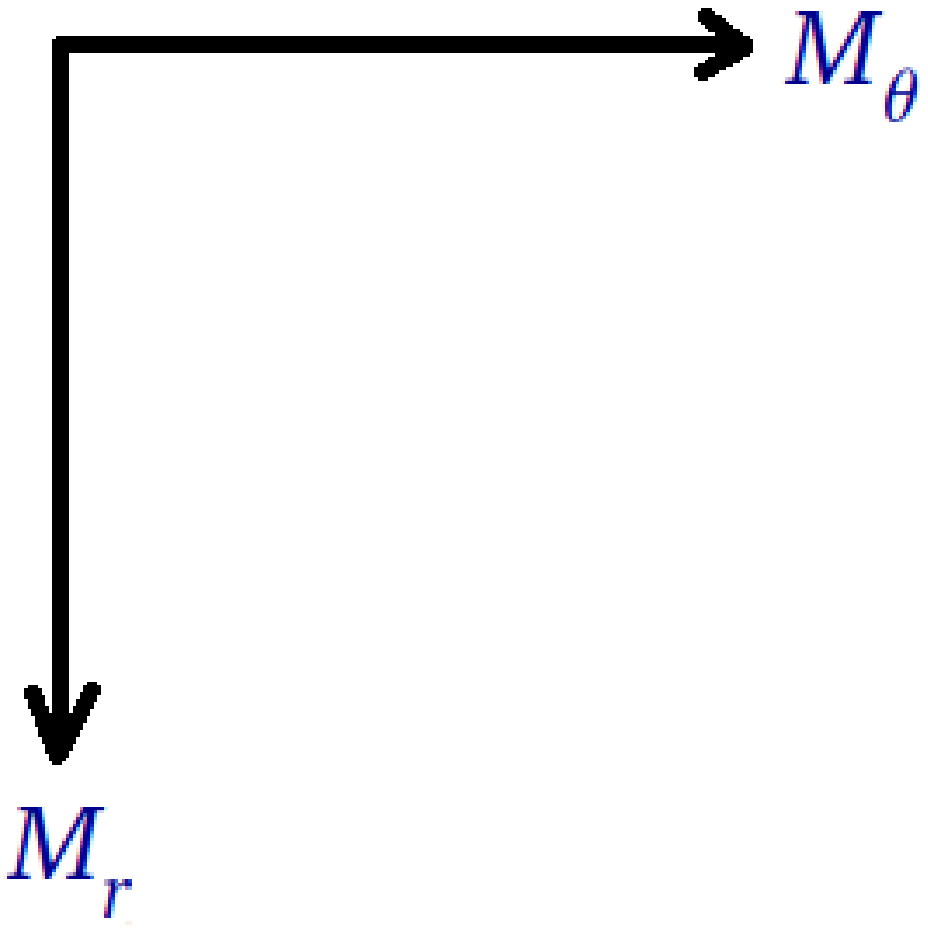}}\hspace{1.8cm}%
\begin{minipage}[c]{0.6\columnwidth}%
\noindent \begin{flushleft}
\includegraphics[scale=0.6]{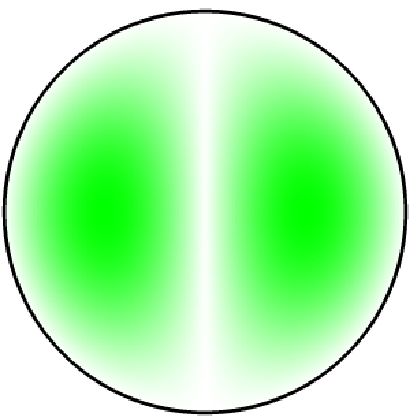}\hspace{.04cm}\includegraphics[scale=0.6]{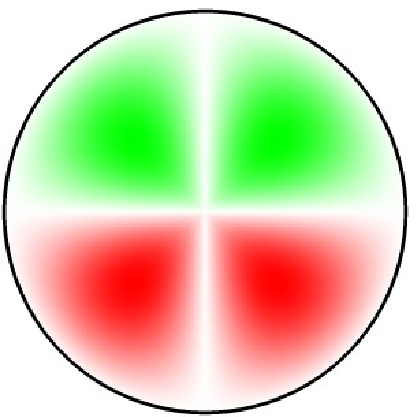}$\put(5,35){\dots}$\\
\includegraphics[scale=0.6]{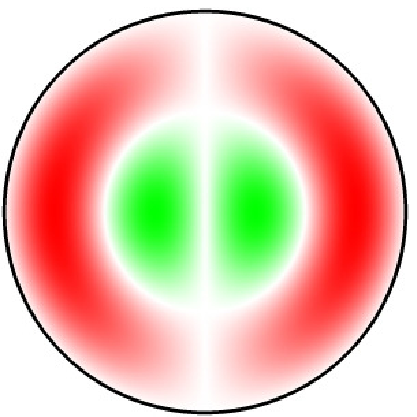}\hspace{.04cm}\includegraphics[scale=0.6]{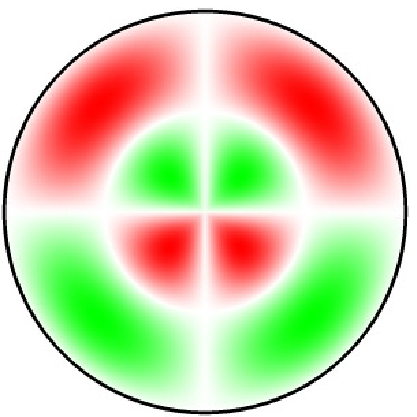}$\put(-107,-15){\vdots}$
\par\end{flushleft}%
\end{minipage}\caption{\textbf{Initial guess of basic field $\chi_{0}$}. Each one of the
four panels shows the distribution of the field $\chi_{0}$ inside
of the neutron star for one choice of the pair $\left({\cal M}_{r},{\cal M}_{\theta}\right)$
according to equation (\ref{eq:51}). The top, left panel visualizes
the case $\left({\cal M}_{r},{\cal M}_{\theta}\right)=\left(1,1\right)$.
Proceeding to the right increases the value of the parameter ${\cal M}_{\theta}$,
and we have to go down to raise the value of ${\cal M}_{r}$. For
each panel, the maximal absolute field value is called $max$. The
values $max$ and $-max$ are represented by the brightest red and
green colors, respectively.\label{fig:Initial}}

\end{figure}

In Sect. \ref{sec:Theory}, we have introduced the basic fields\begin{equation}
\nu,\alpha,\beta,M^{r},M^{\theta},N^{r},N^{\theta},N^{\phi},p,\epsilon,\chi_{0},l_{\phi}\label{eq:50}\end{equation}
and their governing equations (\ref{eq:47}-\ref{eq:8}), (\ref{eq:32}),
(\ref{eq:35}), (\ref{eq:49}) and the equation of state. We solve
these equations by means of a fixed-point iteration after having specified
an initial guess $\nu_{0},...,l_{\phi,0}$, which is constructed in
the following way. We set the basic fields (\ref{eq:50}) to the Tolman-Oppenheimer-Volkoff
solution. This spherically symmetric solution has no meridional circulation,
and therefore the basic matter field $\chi_{0}$ vanishes. Then, we
modify the Tolman-Oppenheimer-Volkoff solution such that the field
$\chi_{0}$ becomes non-zero. For that purpose, we consider surface-adapted
coordinates\begin{eqnarray}
\tilde{r} & = & \frac{r}{R\left(\theta\right)},\label{eq:65}\\
\tilde{\theta} & = & \theta,\nonumber \end{eqnarray}
where $R\left(\theta\right)$ is the radial coordinate of the neutron
star surface. Then, the invariance of scalars tells us\[
\tilde{\chi}_{0}\left(\tilde{r},\tilde{\theta}\right)=\chi_{0}\left(r,\theta\right),\]
and we set\begin{eqnarray}
\tilde{\chi}_{0}\left(\tilde{r},\tilde{\theta}\right) & = & \chi_{0}^{{\rm max}}\sin\left({\cal M}_{r}\tilde{r}\pi\right)\sin\left({\cal M}_{\theta}\tilde{\theta}\right),\;\;\;\;\;\;\;\;\label{eq:51}\end{eqnarray}
with an arbitrary constant $\chi_{0}^{{\rm max}}$ and parameters
${\cal M}_{r},{\cal M}_{\theta}\in\left\{ 1,2,...\right\} $, as shown
in Fig. \ref{fig:Initial}.

To describe the iteration steps, we denote the values of the basic
fields (\ref{eq:50}) after the $s$-th iteration step by $\nu_{s},...,l_{\phi,s}$,
and those after the previous $s$-th iteration step by $\nu_{s-1},...,l_{\phi,s-1}$.
The basic geometry fields are then iterated via the equations\begin{eqnarray}
\nu_{s} & = & ^{3}\Delta^{-1}S_{\nu,s-1},\label{eq:52}\\
\alpha_{s} & = & ^{2}\Delta^{-1}S_{\alpha,s-1}-\nu_{s},\label{eq:53}\\
\beta_{s} & = & \left(^{2}\Delta^{-1}S_{\beta,s-1}\right)/\left(r\sin\theta\right)-\nu_{s},\label{eq:54}\\
M_{s}^{m} & = & e^{-2\left(\alpha_{s}+\nu_{s}\right)}\,{^{2}\Delta_{\;\;\; n}^{m}}^{-1}S_{M,s-1}^{n},\label{eq:55}\\
N_{s}^{a} & = & {^{3}\Delta_{\;\; b}^{a}}^{-1}S_{N,s-1}^{b},\label{eq:56}\end{eqnarray}
where $S_{...,s-1}^{...}$ denotes the sources of equations (\ref{eq:47}-\ref{eq:8})
computed from the basic field values $\nu_{s-1},...,l_{\phi,s-1}$.
The inverse Laplacians $^{...}\Delta_{...}^{...-1}$ appearing above
are determined with Green functions according to the detailed description
given further below (Sect. \ref{sub:Green-functions}). Subsequently
we evaluate the source of equation (\ref{eq:35}) using the newly
computed values $\nu_{s},...,M_{s}^{\theta}$ of the basic geometry
fields and the old values $p_{s-1},...,l_{\phi,s-1}$ of the basic
matter fields, the outcome being called $S_{\chi_{0},s-1}$. Next,
we calculate\begin{eqnarray}
p_{s} & = & H^{-1}\left[H\left(p_{{\rm c}}\right)+\left(\ln\left(u_{t,s-1}^{{\rm c}}\right)^{2}-\ln\left(u_{t,s-1}\right)^{2}\right)/2\right.\nonumber \\
 &  & \;\;\;\;\;\;\;\;\;\left.+I(\psi_{{\rm c}})-I(\psi_{s-1})\right],\nonumber \\
\epsilon_{s} & = & \epsilon\left(p_{s}\right),\nonumber \\
\chi_{0} & = & \left(^{3}\Delta^{-1}\left(S_{\chi_{0},s-1}\cos\phi\right)\right)/\cos\phi,\label{eq:57}\\
l_{\phi,s} & = & L\left(\psi_{s-1}\right),\nonumber \end{eqnarray}
where $u_{t,s-1}^{{\rm c}}$ is the value of the quantity $u_{t}^{{\rm c}}$
computed from the values of the basic fields $\nu_{s-1},...,l_{\phi,s-1}$.

\subsection{Green functions\label{sub:Green-functions}}

In the following, we address the inverse Laplacians $^{2}\Delta^{-1}$,
$^{3}\Delta^{-1}$, ${^{2}\Delta_{\;\;\; n}^{m}}^{-1}$ and ${^{3}\Delta_{\;\; b}^{a}}^{-1}$
appearing in equations (\ref{eq:52}-\ref{eq:57}). Each of these
inverse Laplacians comes from an equation having one of the forms\begin{eqnarray}
^{2}\Delta\Phi & = & S,\label{eq:58}\\
^{3}\Delta\Phi & = & S,\label{eq:59}\\
^{2}\Delta_{\;\;\; n}^{m}\Phi^{n} & = & S^{n},\label{eq:60}\\
^{3}\Delta_{\;\;\; b}^{a}\Phi^{b} & = & S^{a},\label{eq:61}\end{eqnarray}
where $\Phi^{...}$ and $S^{...}$ are a potential and a source, respectively.
To solve for the potential of equation (\ref{eq:58}), (\ref{eq:59}),
..., we have to compute the first, second, ... of the following integrals,
shown in the same order as above \cite{1989.Komatsu}:\begin{eqnarray}
\Phi\left(\vec{x}\right) & = & \int{\rm d}^{2}x'\frac{S\left(\vec{x}'\right)}{2\pi/\ln\left|\vec{x}-\vec{x}'\right|},\label{eq:62}\\
\Phi\left(\vec{x}\right) & = & -\int{\rm d}^{3}x'\frac{S\left(\vec{x}'\right)}{4\pi\left|\vec{x}-\vec{x}'\right|}+...,\label{eq:63}\\
\Phi^{m}\left(r,\theta\right) & = & \frac{\partial\left(r,\theta\right)^{m}}{\partial\left(x,z\right)^{n}}\int{\rm d}^{2}x'\frac{\partial\left(x',z'\right)^{n}}{\partial\left(r',\theta'\right)^{o}}\frac{S^{o}\left(r',\theta'\right)}{2\pi/\ln\left|\vec{x}-\vec{x}'\right|},\nonumber \\
\Phi^{a}\left(r,\theta\right) & = & -\frac{\partial\left(r,\theta,\phi\right)^{a}}{\partial\left(x,y,z\right)^{b}}\int{\rm d}^{3}x'\frac{\partial\left(x',y,'z'\right)^{b}}{\partial\left(r',\theta',\phi'\right)^{c}}\frac{S^{c}\left(r',\theta'\right)}{4\pi\left|\vec{x}-\vec{x}'\right|},\nonumber \end{eqnarray}
where\begin{equation}
\vec{x}=\left(x,z\right)=\left(r\sin\theta,r\cos\theta\right)\label{eq:69}\end{equation}
for the $2$-dimensional and\[
\vec{x}=\left(x,y,z\right)=(r\sin\theta\cos\phi,r\sin\theta\sin\phi,r\cos\theta)\]
for the $3$-dimensional integrals (analogous for $\vec{x}'$) together
with the Jacobian determinants $\partial\left(...\right)^{a}/\partial\left(...\right)^{b}$.
The dots appearing in equation (\ref{eq:63}) represent additional
terms required for the boundary condition of the basic matter field
$\chi_{0}$ and are given in Appendix \ref{sec:Integral-expansion}.
In that Appendix, the main focus is on an expansion of the above integrals
in terms of trigonometric functions and Legendre polynomials. Typically,
only a few terms of the expansions are needed, and neglecting the
remaining ones leads to a strong reduction of the computational cost
necessary to evaluate the integrals.

\subsection{Slicing conditions}

According to Gourgoulhon \textit{et al.~}\cite{1993.Gourgoulhon},
the basic geometry fields $M^{m}$ and $N^{a}$ satisfy the following
slicing conditions\begin{eqnarray*}
\left(N^{2}M^{m}\right)_{||m} & = & 0,\\
N_{\;\;\,|a}^{a} & = & 0,\end{eqnarray*}
where the $2$-and $3$-covariant derivatives `$||$' and `$|$' are
defined in Appendix \ref{sec:Flat-space-Laplacians}. A lengthy, but
elementary calculation shows that these two slicing conditions can
be rewritten as the vanishing flat-space $2$- and $3$-dimensional
derivatives\begin{eqnarray*}
\partial_{m}M_{e}^{m}+\frac{1}{r}M_{e}^{r} & = & 0,\\
\partial_{m}N_{e}^{m}+\frac{2}{r}N_{e}^{r}+\cot\theta N_{e}^{\theta} & = & 0,\end{eqnarray*}
of the quantities\begin{eqnarray*}
M_{e}^{m} & = & e^{2\left(\alpha+\nu\right)}M^{m},\\
N_{e}^{a} & = & e^{2\alpha+\beta}N^{a}.\end{eqnarray*}
Contrary to an analytic one, a numeric evaluation of the Green functions
of Sect. \ref{sub:Green-functions} always produces values for the
fields $M^{m}$ and $N^{a}$ which somewhat violate the slicing conditions.
Therefore, we consider a $2$- and $3$-dimensional Helmholtz decomposition
and set the gradient parts of the two quantities $M_{e}^{m}$ and
$N_{e}^{a}$ equal to zero such that the divergences of the remaining
parts vanish as demanded by the slicing conditions. Such a procedure
is valid, because when the fixed point iteration has converged, the
gradient parts (which have to be set to zero) decrease with increasing
numerical accuracy, i.e. with increasing grid resolution and number
of trigonometric functions and Legendre polynomials used in the integral
expansions of the Green functions.

\subsection{Final gauge}

Even now, the basic fields are not yet determined completely. There
are still remaining gauge degrees of freedom left, because we can
add an arbitrary constant to each of the potentials appearing in equations
(\ref{eq:58}-\ref{eq:61}) without violating these equations:\begin{equation}
\Phi^{...}\rightarrow\Phi^{...}+{\rm const}^{...}.\label{eq:72}\end{equation}
For the $3$-dimensional Poisson equations, the Green function\[
^{3}G\left(\vec{x},\vec{x}'\right)=-\frac{1}{4\pi\left|\vec{x}-\vec{x}'\right|}\]
vanishes in the limit $\left|\vec{x}-\vec{x}'\right|\rightarrow\infty$.
Thus, we choose the potentials of these Poisson equations to vanish
at infinity by setting the constant appearing in equation (\ref{eq:72})
equal to zero. However, for the $2$-dimensional Poisson equations
such a choice is not possible, as the Green function\[
^{2}G\left(\vec{x},\vec{x}'\right)=\frac{1}{2\pi}\ln\left|\vec{x}-\vec{x}'\right|\]
is not bounded for $\left|\vec{x}-\vec{x}'\right|\rightarrow\infty$.
Therefore, we proceed as follows:

The potential $r\sin\theta\left(\beta+\nu\right)$ of the $2$-dimensional
Poisson equation (\ref{eq:54}) obeys the Dirichlet boundary condition
according to Appendix \ref{sec:Integral-expansion}. Hence, we set
the constant in equation (\ref{eq:72}) equal to zero in this case,
which completely fixes the field $\beta$. In case of the $2$-dimensional
Poisson equation (\ref{eq:53}), we choose the constant in equation
(\ref{eq:72}) in such a manner that\[
\alpha\left(r=0\right)=\beta\left(r=0\right).\]
For the remaining $2$-dimensional Poisson equation (\ref{eq:55}),
we have to consider only the Cartesian $z$-component $M_{e}^{z}$,
because the $x$-component $M_{e}^{x}$ obeys the Dirichlet boundary
condition, and we impose the condition\[
M_{e}^{z}\left(r=0\right)=0.\]

\subsection{Circulation modes}

Eventually, we have a closer look at the function $f\left(\psi\right)$
introduced in equation (\ref{eq:34}). We follow the same approach
as Eriguchi \textit{et al.~}\cite{1986.Ewald} and limit ourselves
to the power law\begin{equation}
f\left(\psi\right)=k\psi^{n},\label{eq:74}\end{equation}
with some constant $k$ and the exponent\begin{equation}
n=0,1.\label{eq:77}\end{equation}
For the case\[
f\left(\psi\right)=k\psi,\]
we find different meridional circulation modes $\psi_{m}$ (with $m=0,1,2,...\,$),
as in the Newtonian case \cite{1986.Ewald}. The distributions of
the field $\chi_{0}$ belonging to these modes look similar to the
ones displayed in Fig. \ref{fig:Initial} but are somewhat deformed.

Unfortunately, when nearly having reached one of the higher circulation
modes $\psi_{1},\psi_{2},...$ during the fixed point iteration, the
fixed point iteration method always starts to converge to the fundamental
mode $\psi_{0}$. In order to obtain the higher modes, we therefore
projected the lower ones away. For that purpose, we assume that we
have already evaluated the first $m_{{\rm e}}-1$ modes, i.e. we know
$\psi_{m}$ for $m=0,1,...,m_{{\rm e}}-1$. Then, the fixed point
iteration leads to the $m_{{\rm e}}$-th mode by replacing\[
\psi\rightarrow\psi-\sum_{m=1}^{m_{{\rm e}}-1}C_{m}\psi_{m}\]
at every iteration step, with adequately chosen coefficients $C_{m}$.
If an orthogonality relation\begin{equation}
\int_{0}^{\infty}{\rm d}r\int_{0}^{\pi}{\rm d}\theta\sqrt{h}W_{m}\left(r,\theta\right)\psi_{m}\psi_{m'}=\delta_{mm'}\label{eq:78}\end{equation}
with the weight function $W_{m}\left(r,\theta\right)$ exists, the
coefficients $C_{m}$ are given by\[
C_{m}=\frac{\int_{0}^{\infty}{\rm d}r\int_{0}^{\pi}{\rm d}\theta\sqrt{h}W_{m}\left(r,\theta\right)\psi\psi_{m}}{\int_{0}^{\infty}{\rm d}r\int_{0}^{\pi}{\rm d}\theta\sqrt{h}W_{m}\left(r,\theta\right)\psi_{m}\psi_{m}},\]
where $h={\rm det}h_{ab}$. However, we neither know whether an orthogonality
relation exists nor do we know the weight functions $W_{m}\left(r,\theta\right)$.
After some experimenting, we found that the choice\begin{equation}
W_{m}\left(r,\theta\right)=\epsilon\label{eq:73}\end{equation}
is sufficient to achieve a convergence to higher modes. This does
not necessarily mean that (\ref{eq:73}) is the correct weight function,
but it could be very close to it.

In addition to the usage of (\ref{eq:73}), we perform the following
steps in the GRNS code: The pressure distribution of the solutions
investigated in this work is always equatorially symmetric. However,
in our treatment, equatorial symmetry is not guaranteed exactly due
to the limited numerical accuracy. Therefore, equatorial asymmetry
may increase during the fixed point iteration, eventually leading
to an undesired meridional circulation mode. In order to prevent this,
we symmetrize the pressure distribution at every iteration step. A
similar method is performed for the basic field $\chi_{0}$, which
has either even or odd parity depending on the considered mode.

\subsection{GRNS}

We have implemented GRNS under Linux in C++, and it possesses an OpenGL
visualization interface, which allows the user to supervise the fixed
point iteration. The user has full control over the iteration, which
can be stopped and restarted at any time. The user can select any
of the physical fields either when the iteration is stopped or even
when it is running, and display it on the screen. It is also possible
to visualize the flow of the neutron star fluid in real time and to
see how the neutron star surface changes during the iteration. These
code features are of advantage when analyzing the stability of GRNS,
and they were also very helpful in debugging the code.

We used a numerical grid of 150 radial and 156 angular zones to compute
the models discussed in Sect. \ref{sec:Results}. For the sums arising
from the expansion of the integrals (see Appendix \ref{sec:Integral-expansion}),
we chose the upper limit\begin{equation}
\sum_{l=...}^{\infty}\rightarrow\sum_{l=...}^{l_{{\rm max}}},\label{eq:80}\end{equation}
where $l_{{\rm max}}=10$ for all models of Sect. \ref{sec:Results}.
This limits the computational cost, and we are able to compute individual
circulation modes within about a minute on a current single core CPU.
In Appendix \ref{sec:Tests}, we discuss the convergence behavior
and consistency tests of GRNS.

\section{Results\label{sec:Results}}

\subsection{Reference model}

The results presented in the following were obtained with the usual
polytropic equation of state\begin{eqnarray*}
p & = & K\rho^{\Gamma},\\
\epsilon & = & \rho+\frac{p}{\Gamma-1},\end{eqnarray*}
where $K$ is the polytropic constant, $\rho$ the rest mass density,
and $\Gamma$ the polytropic exponent, respectively. From this equation
of state its is readily seen that\begin{equation}
\epsilon=\left(\frac{p}{K}\right)^{\frac{1}{\Gamma}}+\frac{p}{\Gamma-1}.\label{eq:75}\end{equation}
We keep the maximum absolute value of the stream function $\psi$
fixed at the value $\psi_{{\rm max}}$ such that it does not matter
which value we choose for the constant $k$ in equation (\ref{eq:74}).
The fundamental mode of each neutron star model is then unambiguously
specified by the following parameters\[
n,p_{{\rm c}},\rho_{{\rm c}},\Gamma,\psi_{{\rm max}},L\left(\psi\right)\]
i.e. by the exponent appearing in equation (\ref{eq:74}), the central
pressure and the central density (which fixes the polytropic constant
$K$) of the initial guess, the polytropic exponent, the maximum absolute
value of the stream function, and by the distribution (\ref{eq:32})
of the specific angular momentum. As a reference model, we choose
a non-rotating neutron star with parameters\begin{eqnarray*}
n & = & 1,\\
p_{{\rm c}} & = & 9.1\cdot10^{34}\,{\rm erg}/{\rm cm}^{3},\\
\rho_{c} & = & 7.9\cdot10^{14}\,{\rm g}/{\rm cm}^{3},\\
\Gamma & = & 2,\\
\psi_{{\rm max}} & = & 3\cdot10^{34}\,{\rm g}/{\rm s},\\
L\left(\psi\right) & = & 0,\end{eqnarray*}
circulating at its fundamental mode. This set of parameters corresponds
to a neutron star with the following properties:\begin{eqnarray}
\bar{R} & = & 15.6\,{\rm km},\label{eq:76}\\
{\cal M} & = & 1.51\, M_{\odot},\nonumber \\
\bar{v}_{{\rm circ}} & = & 1043\,{\rm km}/{\rm s}.\nonumber \end{eqnarray}
The average radius $\bar{R}$ is the average proper radius\[
R_{C}=\int_{C}{\rm d}re^{\alpha}\]
averaged over all angles $\theta$ (over all radial paths $C$). We
compute the rest mass via \cite{1992.Cook}\[
{\cal M}=\int_{V}{\rm d}r{\rm d}\theta{\rm d}\phi\sqrt{h}Nu^{t}\rho,\]
where $V$ is the volume of the star and the average circulation velocity
is given by\[
\bar{v}_{{\rm circ}}=\left\langle \sqrt{\left(v^{r}\right)^{2}+\left(rv^{\theta}\right)^{2}}\right\rangle \]

\begin{figure}
\noindent \centering{}\hspace{-3.8cm}\includegraphics[scale=0.3]{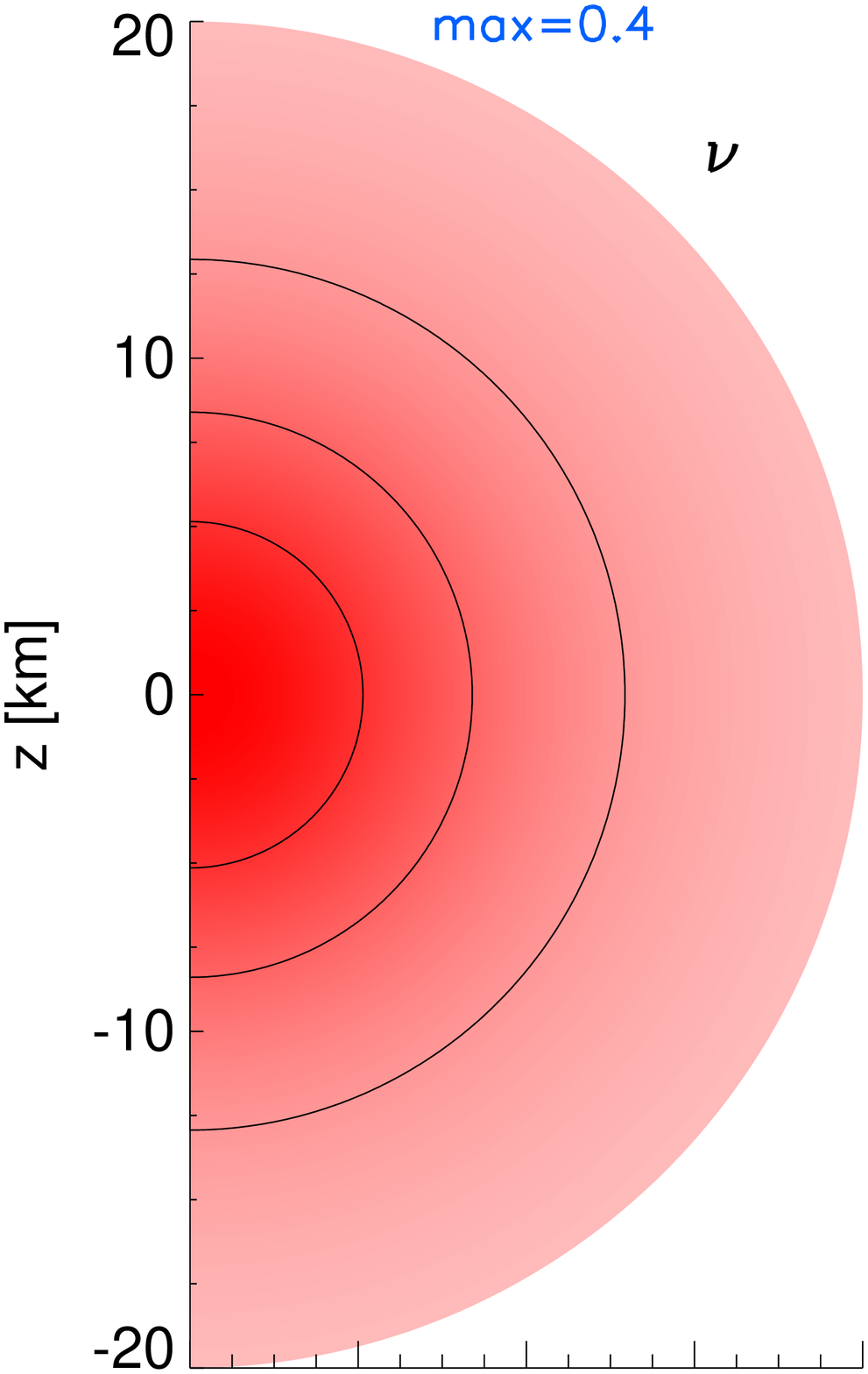}\put(-62.5,0){\includegraphics[scale=0.3]{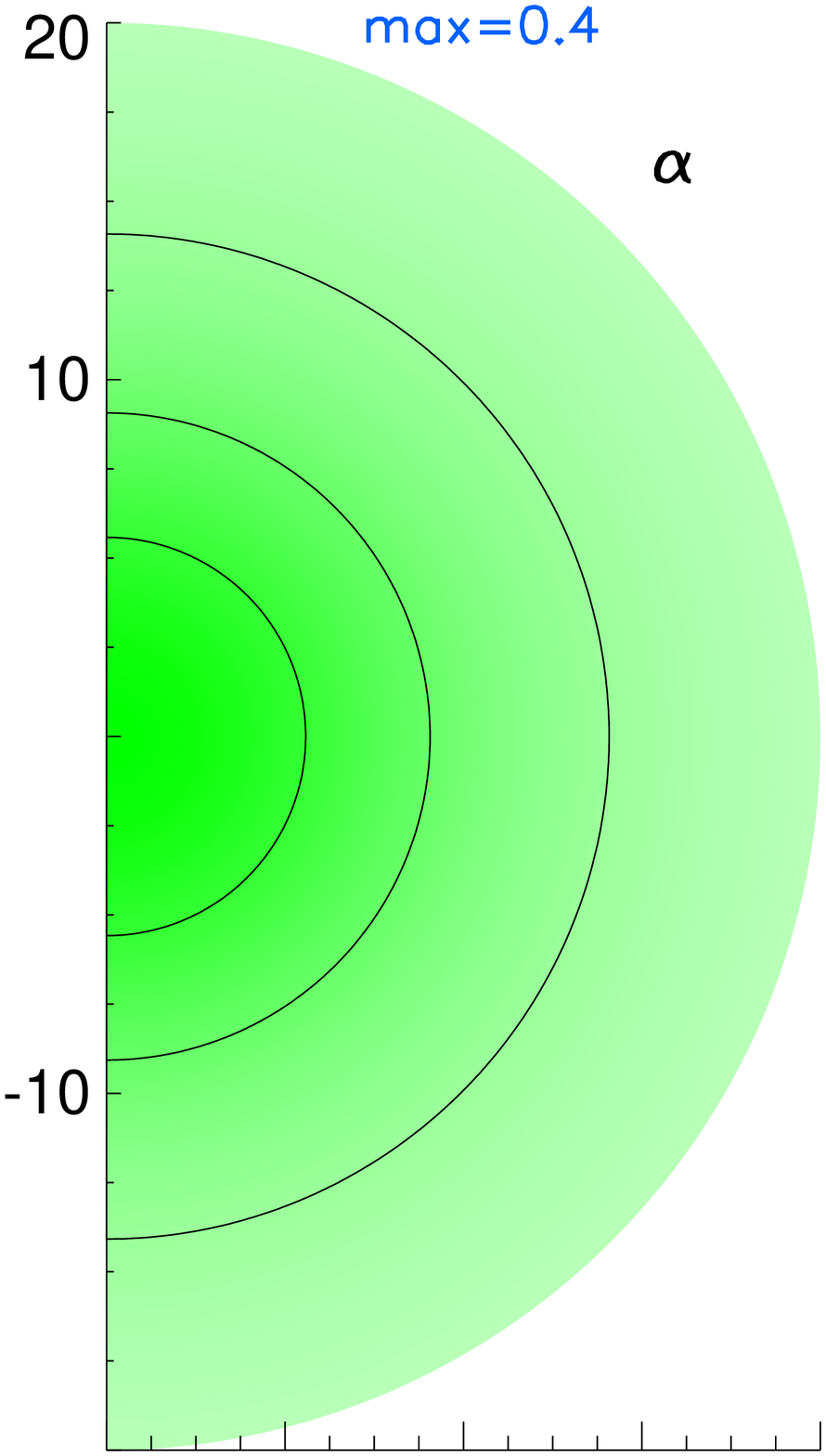}}\\
\vspace{-1.2cm}\hspace{-3.8cm}\includegraphics[scale=0.3]{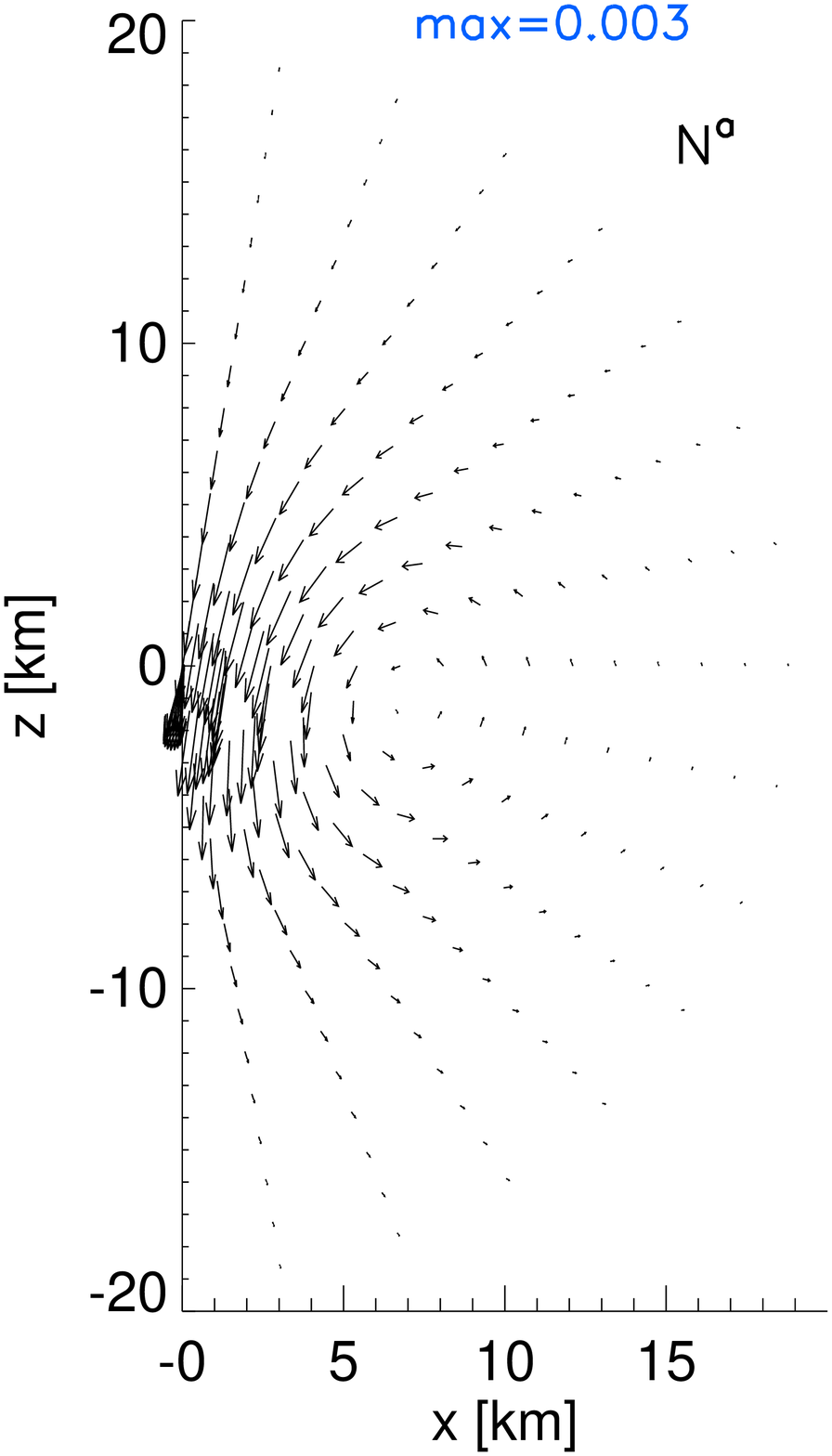}\put(-62.5,0){\includegraphics[scale=0.3]{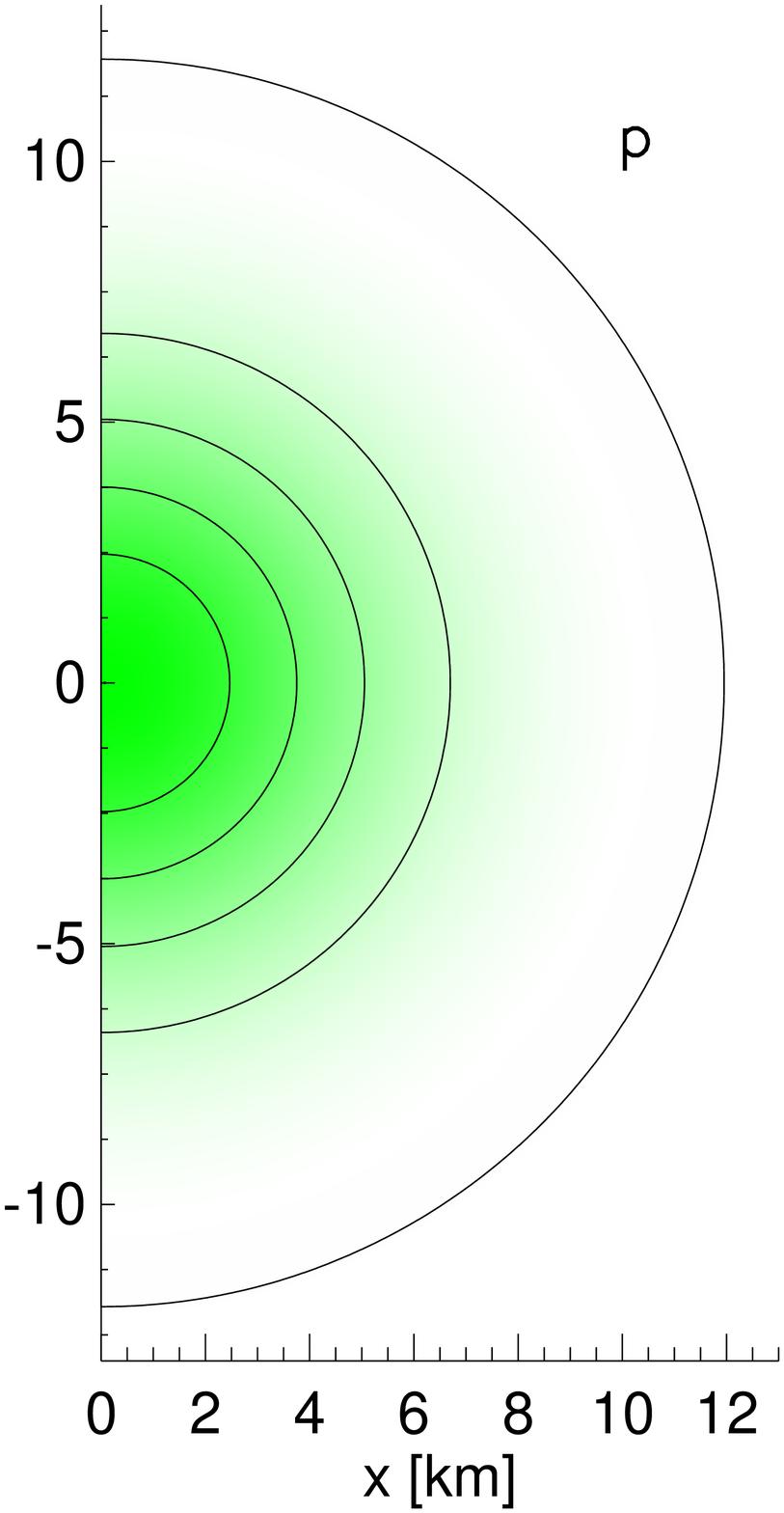}}\caption{\textbf{Basic fields of reference model}. The four panels show the
$3$-lapse logarithm $\nu$, the $2$-conformal factor logarithm $\alpha$,
the $3$-shift $N^{a}$ and the pressure for our reference model.
The color coding of the three scalar plots is the same one as in Fig.
\ref{fig:Initial}, where the maximum absolute field value $max$
is displayed at the top of each panel. For the vector plot, the quantity
$max$ denotes the maximum vector length (note that all vectors lie
within the paper plane).\label{fig:Reference}}

\end{figure}

\begin{figure}
\noindent \centering{}\hspace{-3.8cm}\includegraphics[scale=0.3]{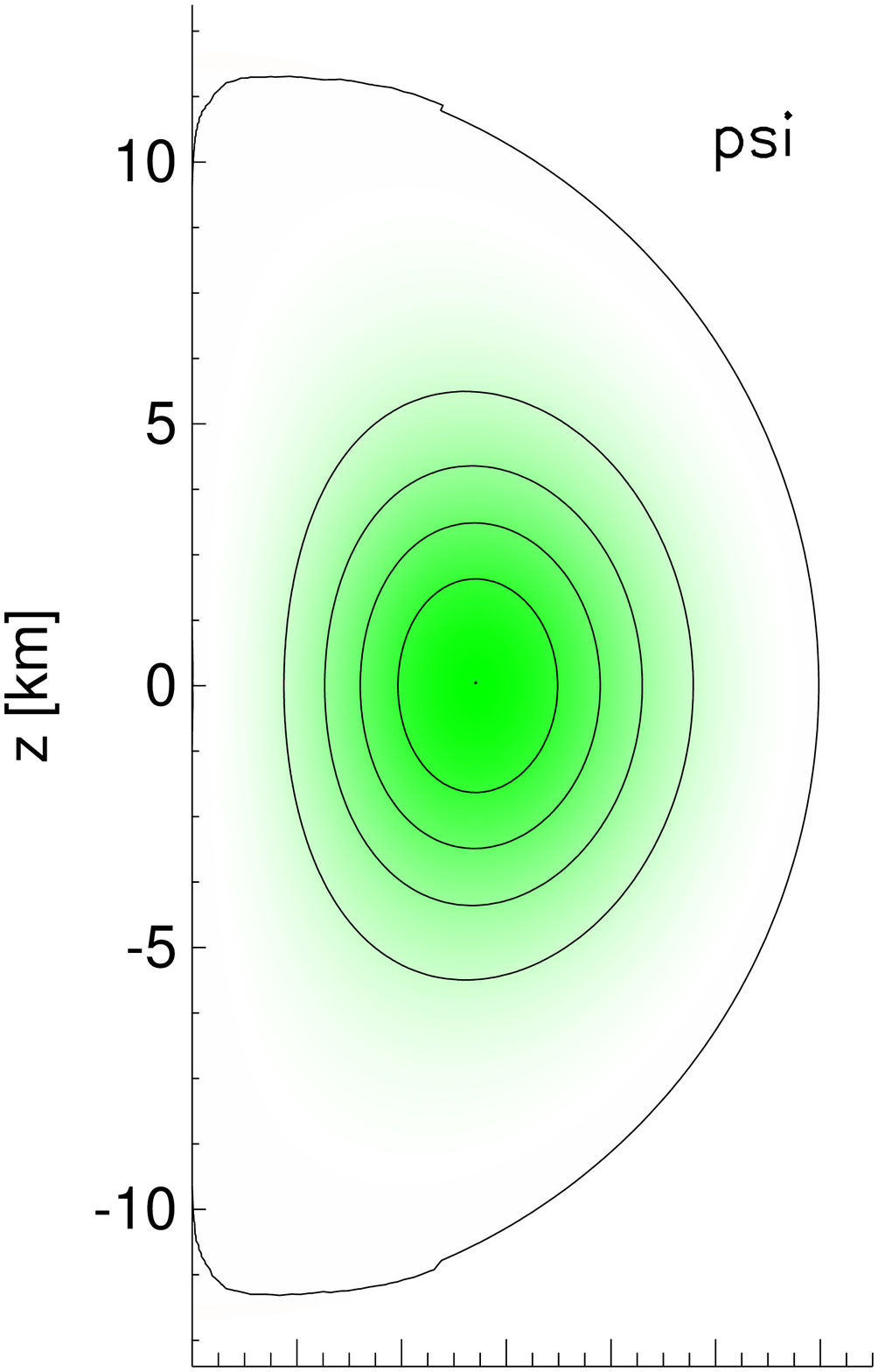}\put(-62.5,0){\includegraphics[scale=0.3]{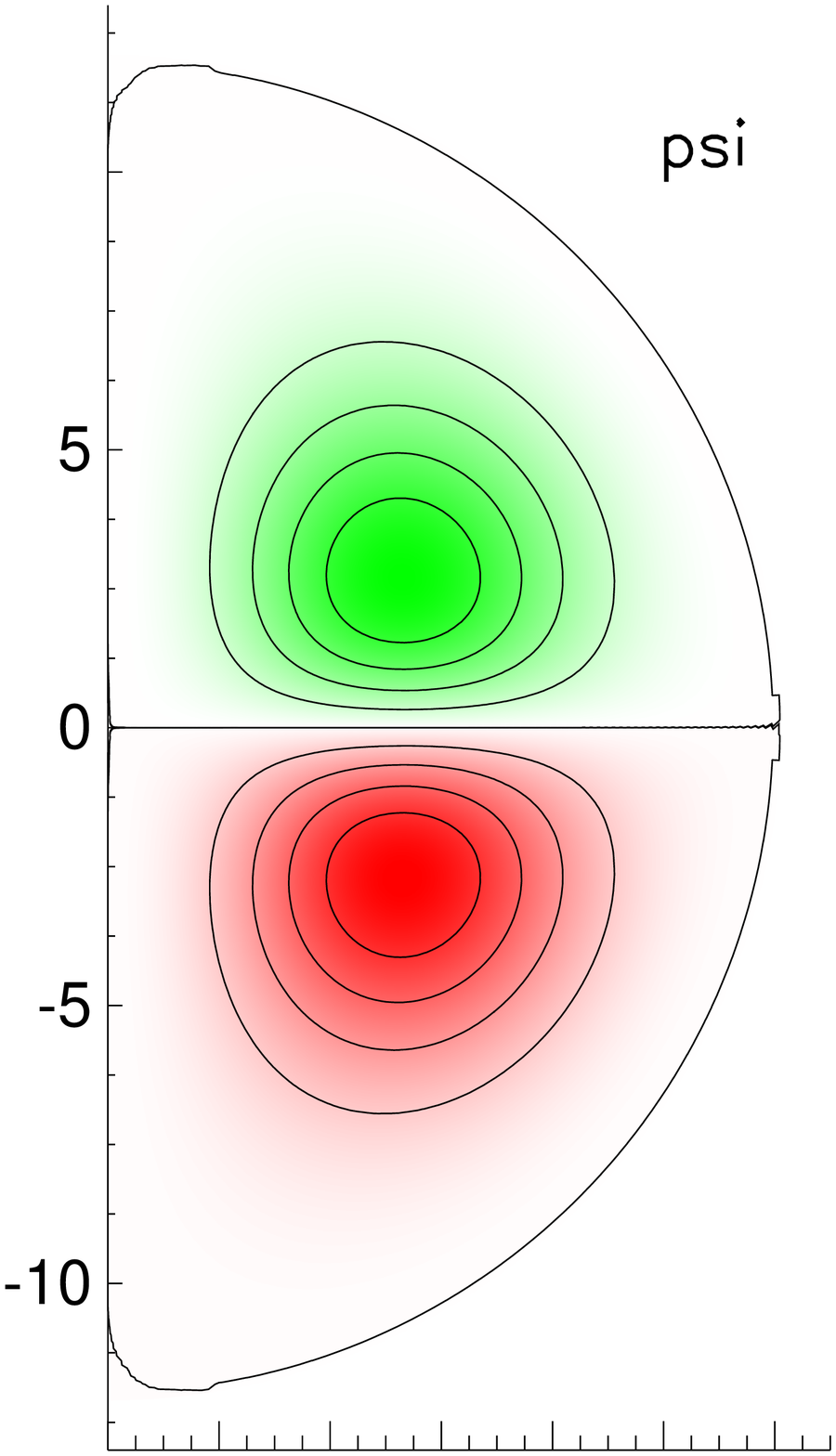}}\\
\vspace{-1.53cm}\hspace{-3.8cm}\includegraphics[scale=0.3]{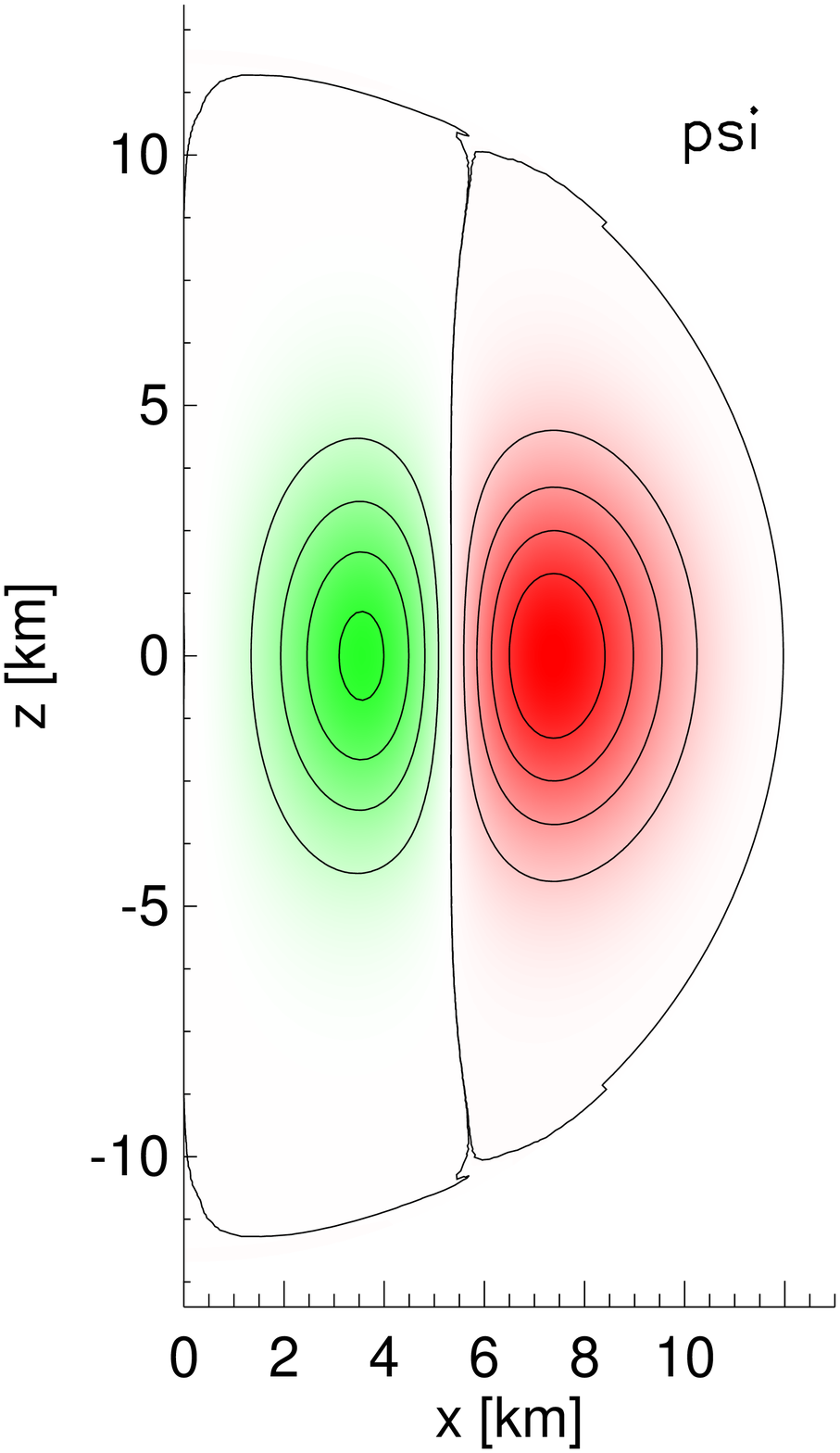}\put(-62.5,0){\includegraphics[scale=0.3]{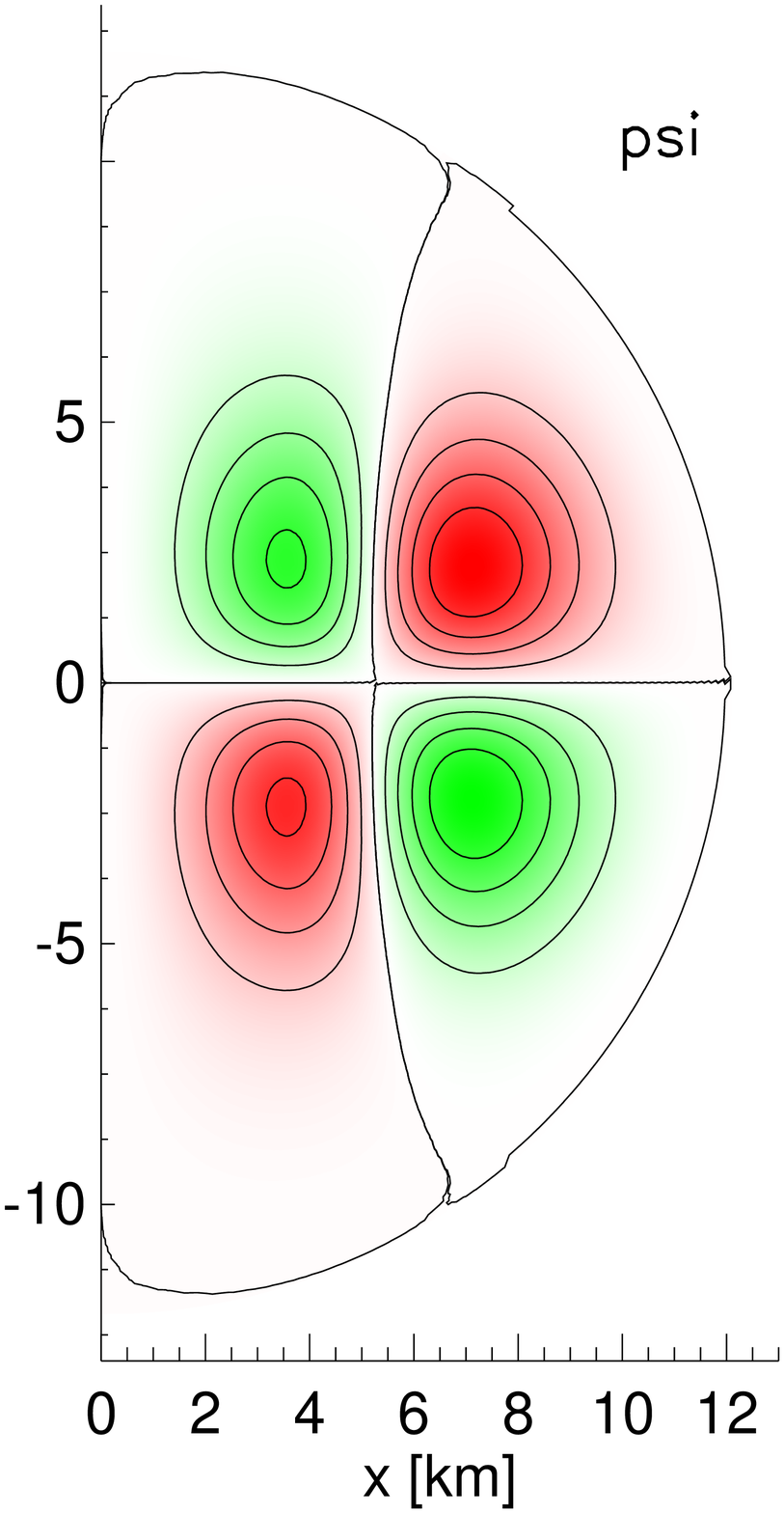}}\caption{\textbf{Circulation modes - Part 1}. The panels show the stream lines
of the first four meridional circulation modes of the reference model.
The color coding is the same as in Fig. \ref{fig:Initial}. Note that
red and green refer to clockwise and counterclockwise motion, respectively.
The kinks visible at the neutron star surface are a result of the
finite numerical resolution and the use of contour plots.\label{fig:Modes-1}}

\end{figure}

The basic fields of the reference model are shown in Fig. \ref{fig:Reference}.
The $3$-lapse logarithm $\nu$ is one of the two contributions (the
other one comes from the 3-shift $N^{a}$) to the time dilation, which
is strongest at the center of the neutron star. The $2$-conformal
factor logarithm $\alpha$ represents the stretching of space in meridional
planes and is also strongest in the center. We do not display the
basic geometry field $\beta$ in Fig. \ref{fig:Reference}, because
its values are not significantly different from those of the field
$\alpha$. The $3$-shift $N^{a}$ represents the dragging of spacetime
caused by the flow of the neutron star fluid. Since the reference
model is circulating, this leads to the vortex visible in the lower
left panel of Fig. \ref{fig:Reference}. The total energy density
$\epsilon$ is also not shown in that figure, because it is correlated
to the pressure $p$ by the trivial analytic relation (\ref{eq:75}).
Note that in the lower right panel of Fig. \ref{fig:Reference} the
outermost contour is the surface of the neutron star. As the values
on the axes of Fig. \ref{fig:Reference} do not refer to proper but
to coordinate distances, the average neutron star radius suggested
by the contour is somewhat smaller than the actually correct value
(\ref{eq:76}), which takes the curvature of spacetime into account.
The $2$-shift $M^{m}$ and specific angular momentum $l_{\phi}$
vanish for the reference model, because it does not rotate. Instead
of considering the modified stream function $\chi_{0}=\psi/\left(r\sin\theta\right)$,
we show the stream function $\psi$ itself in the top left panel of
Fig. \ref{fig:Modes-1}, as the contours of the stream function $\psi$
are stream lines of the meridional flow. Note that the stream function
of the reference model contains only a single vortex.

The distributions of the $3$-lapse $\nu$, the $2$-conformal factor
$\alpha$ and the pressure $p$ given in Fig. \ref{fig:Reference}
do not show any significant difference when compared to the Tolman-Oppenheimer-Volkoff
solution used to construct the initial guess for the fixed point iteration.
The only relevant changes are that the $3$-shift $N^{a}$ and the
stream function $\psi$, which vanish for the Tolman-Oppenheimer-Volkoff
solution, display the distributions shown in the lower, left panel
of Fig. \ref{fig:Reference} and the upper left panel of Fig. \ref{fig:Modes-1},
respectively.

\subsection{Higher circulation modes\label{sub:Higher-circulation-modes}}

The other three panels in Fig. \ref{fig:Modes-1} show higher circulation
modes with additional vortices. Actually, each vortex in the stream
function $\psi$ is accompanied by a vortex in the $3$-shift $N^{a}$
(not shown for the higher modes). Let us now compare the four modes
shown in Fig. \ref{fig:Modes-1} with the four initial guesses of
Fig. \ref{fig:Initial}. With this in mind, it does not matter that
both figures show different fields, because the difference between
the stream function $\psi$ and the modified stream function $\chi_{0}$
is merely a factor $r\sin\theta$. Such a factor causes deformations
of the vortices, but their number and orientations remain unaffected.
Then, it seems as if the top left initial guess of Fig. \ref{fig:Initial}
leads to the top left mode of Fig. \ref{fig:Modes-1}, and so on.
However, there is no such one-to-one correlation. Instead, it can,
for instance, occur that the four modes of Fig. \ref{fig:Modes-1}
are produced by the four initial guesses ${\cal M}_{r}=1$ and ${\cal M}_{\theta}=1,2,3,4$.
Actually, it does not matter too much how the initial guesses are
chosen. It is only important that they are somehow different, and
to improve the fixed point iteration, we have made a choice that is
at least similar to the expected outcomes.

\begin{figure}
\noindent \centering{}\hspace{-3.8cm}\includegraphics[scale=0.3]{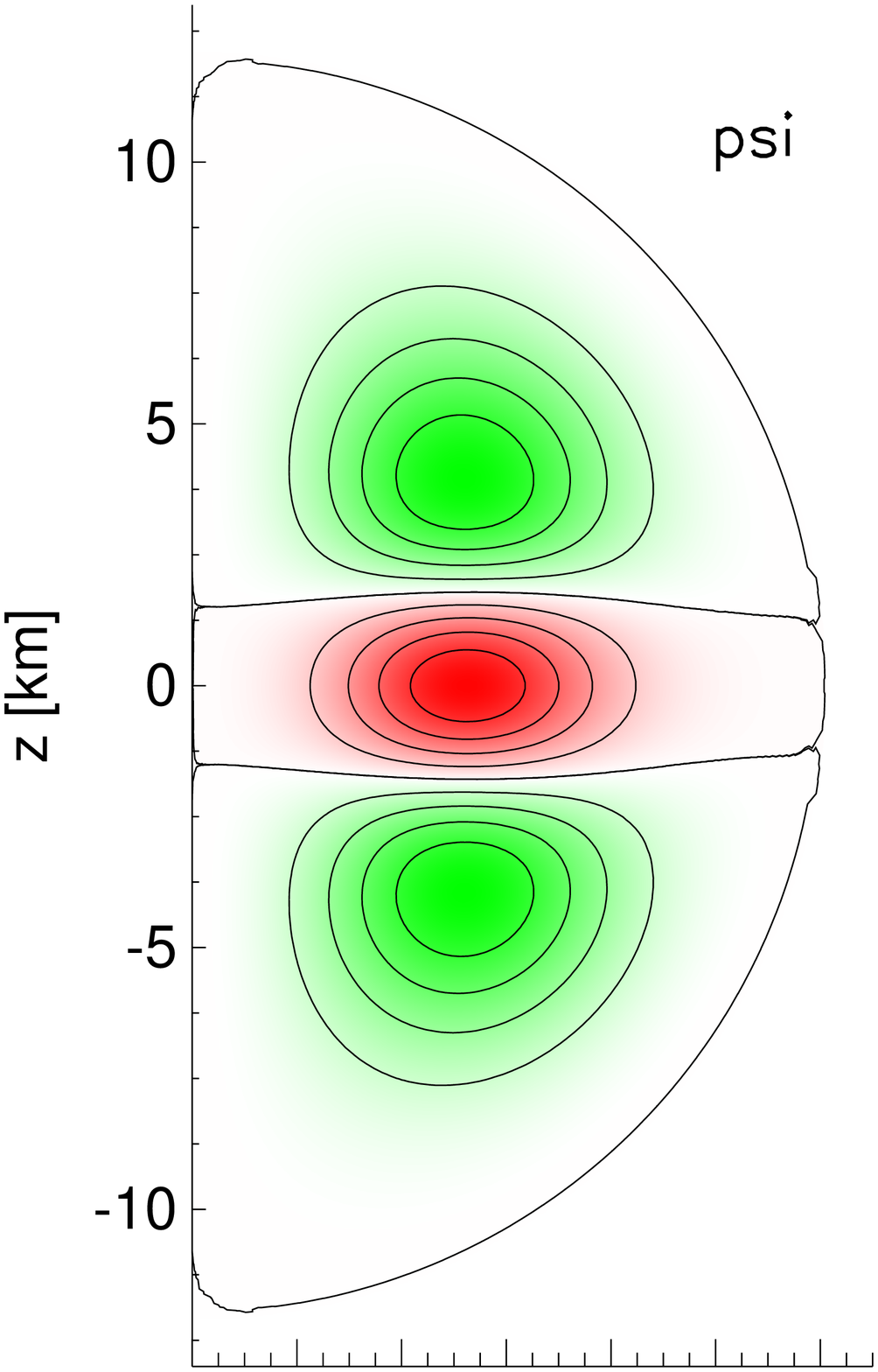}\put(-62.5,0){\includegraphics[scale=0.3]{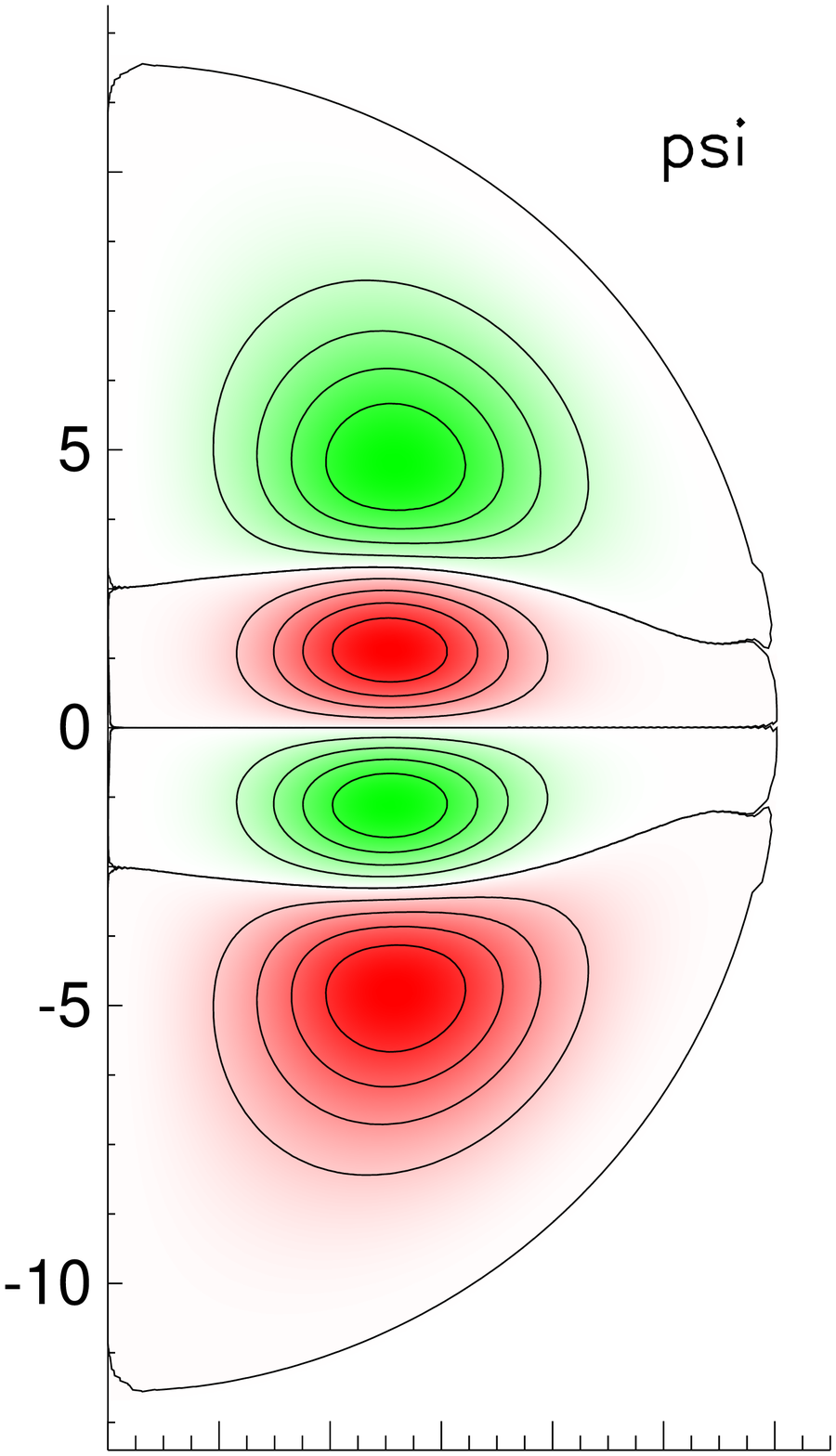}}\\
\vspace{-1.53cm}\hspace{-3.8cm}\includegraphics[scale=0.3]{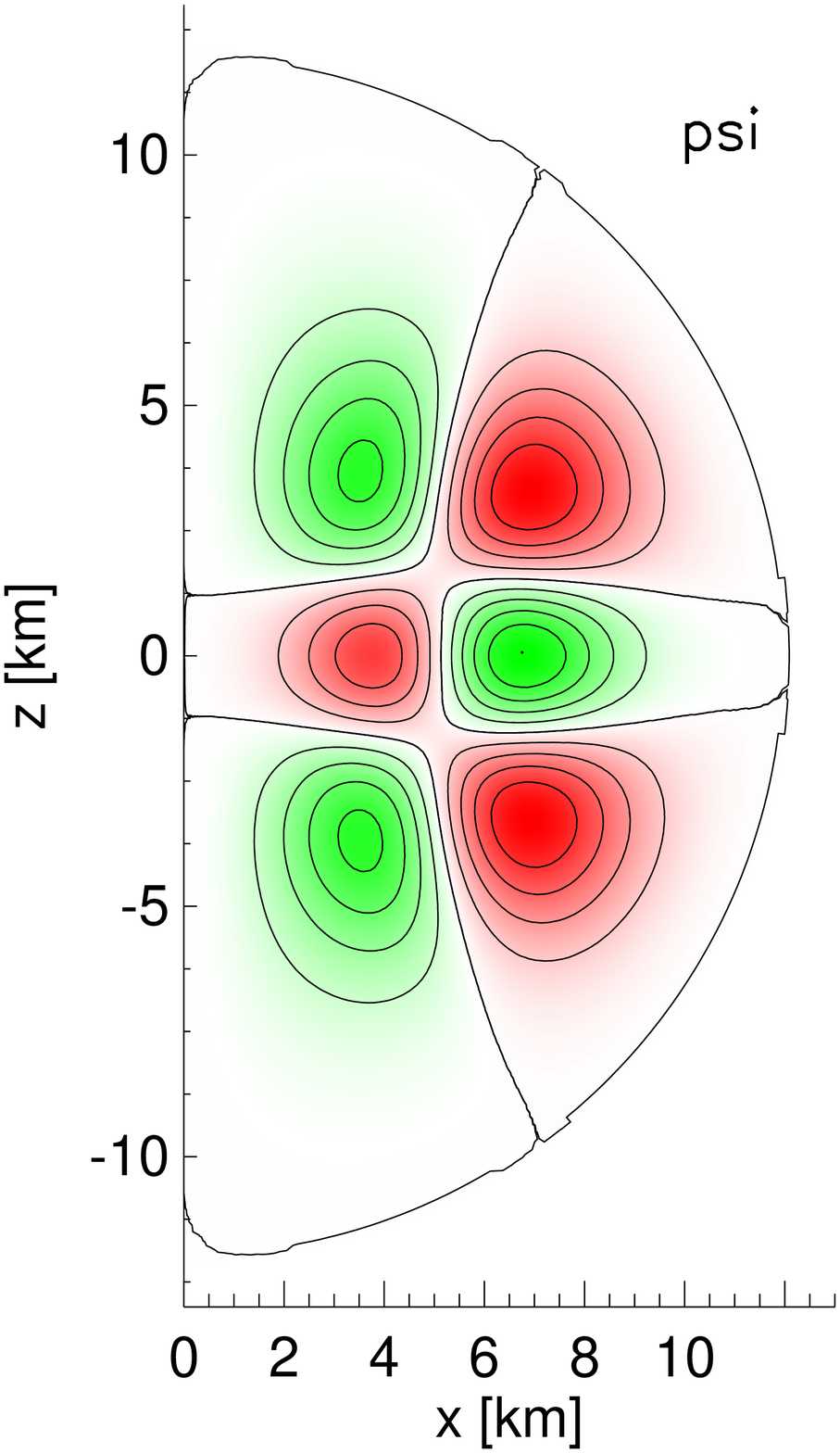}\put(-62.5,0){\includegraphics[scale=0.3]{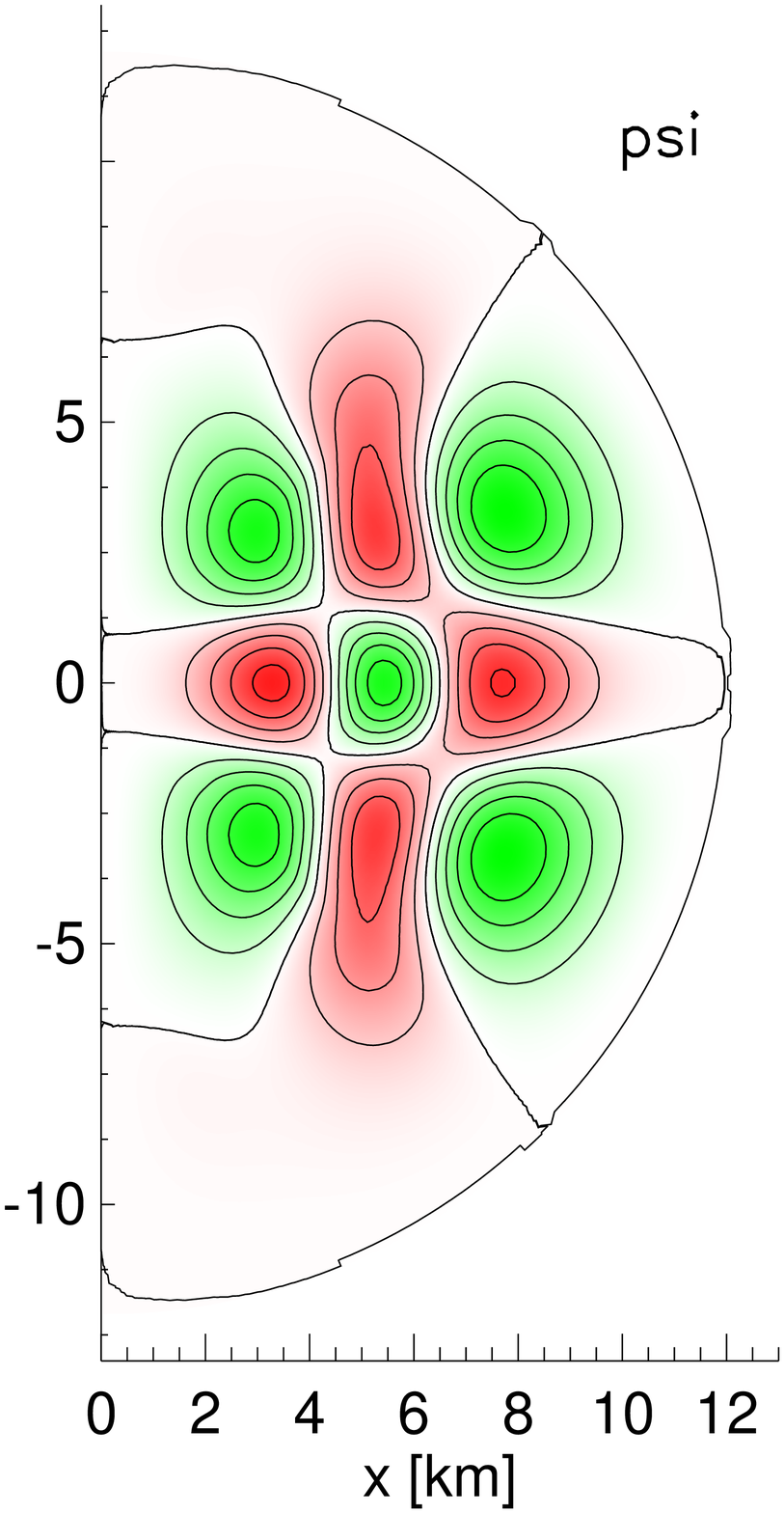}}\caption{\textbf{Circulation modes - Part 2}\label{fig:Modes-2}}

\end{figure}

In total, we are able to compute 16 modes for the reference model
(including the fundamental mode) before GRNS fails due to too large
numerical errors, i.e. we find a much larger set of modes than in
Eriguchi \textit{et al.~}\cite{1986.Ewald}. Fig. \ref{fig:Modes-2}
shows some additional higher modes. The lower right panel of that
figure represents the highest mode obtained where numerical errors
do not yet have a significant impact on the shape of the vortices.
Looking at Figs. \ref{fig:Modes-1} and \ref{fig:Modes-2}, it is
obvious that the circulation modes constitute a $2$-dimensional mode
set. 

\subsection{Surface deformations}

\begin{figure}
\noindent \centering{}\hspace{-.1cm}\includegraphics[scale=0.25]{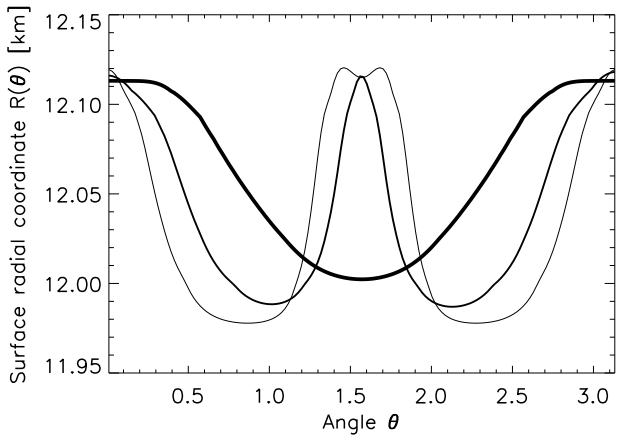}\caption{\textbf{Surface deformations. }The figure shows the radial coordinate
$R\left(\theta\right)$ of the neutron star surface depending on the
angle $\theta$. From thick to thin, the three lines refer to the
upper left and upper right mode of Fig. \ref{fig:Modes-1}, and the
upper left mode of Fig. \ref{fig:Modes-2}.\label{fig:Surface}}

\end{figure}

A closer look at the surface of the neutron star in Fig. \ref{fig:Surface}
reveals that surface deformations of about half a percent of the average
neutron star radius are present. The shape of these deformations depends
on the vortices in the vicinity of the surface. Vortices deeper inside
of the neutron star have only a small impact on the deformations.
In Eriguchi \textit{et al.~}\cite{1986.Ewald}, the numerical resolution
was too small to resolve such properties. However, it was determined
that for the fundamental mode with only a single vortex the meridional
circulation causes the neutron star to become slightly prolate. 

The strength of the deformations depends also on the velocity of the
fluid in the vortices, and thus on the maximum absolute value $\psi_{{\rm max}}$
of the stream function. To test the limits, we increased the value
of $\psi_{{\rm max}}$ in GRNS as far as possible. The choice $\psi_{{\rm max}}=3\cdot10^{34}\,{\rm g}/{\rm s}$
for the reference model is already close to what is possible with
GRNS, and we were only able to increase that value to about $\psi_{{\rm max}}=10^{35}\,{\rm g}/{\rm s}$,
above which numerical errors grow dramatically. However, even at the
limiting value, the surface deformations do not exceed about one percent
of the average neutron star radius.

\subsection{Circulation velocities}

For the choice $\psi_{{\rm max}}=10^{35}\,{\rm g}/{\rm s}$, the average
circulation velocity of the reference model is $\bar{v}_{{\rm circ}}=3431\,{\rm km}/{\rm s}$,
whereas for a very slowly circulating model with $\psi_{{\rm max}}=3\cdot10^{34}\,{\rm g}/{\rm s}$
the average circulation velocity is only $\bar{v}_{{\rm circ}}=10\,{\rm km}/{\rm s}$.
Actually, it is a good approximation to say that the average circulation
velocity $\bar{v}_{{\rm c}}$ scales linearly with the maximum absolute
value $\psi_{{\rm max}}$ of the stream function. However, for the
average radius and the mass of the neutron star, we do not find any
significant changes when the circulation velocity changes, within
the range considered here.

Tab. \ref{tab:Average-circulation-velocities.} lists the average
velocities $\bar{v}_{{\rm circ}}$ of the circulation modes. For that
purpose, we use a mode numbering similar to Fig. \ref{fig:Initial}:
the first mode number in Tab. \ref{tab:Average-circulation-velocities.}
gives the number of vortices in $x$-direction, and the second one
in $z$-direction. The velocity $\bar{v}_{{\rm circ}}$ obviously
increases with the mode numbers, which is a consequence of the orthogonality
relation (\ref{eq:78}). As the average circulation velocity $\bar{v}_{{\rm circ}}$
rises with the maximum absolute value $\psi_{{\rm max}}$ of the stream
function, changing the orthogonality relation (\ref{eq:78}) by inserting
an appropriate factor $D_{m}$ in the integrand, it is possible to
create new modes with the same velocity $\bar{v}_{{\rm circ}}$, but
a different maximum $\psi_{{\rm max}}$. Because it is not easy to
determine the factor $D_{m}$, our method is more practical.

\begin{table}
\caption{\textbf{Average circulation velocities}. The table shows the average
velocities $\bar{v}_{{\rm circ}}$ for the eight circulation modes
of Figs. \ref{fig:Modes-1} and \ref{fig:Modes-2}.\label{tab:Average-circulation-velocities.}}
\begin{tabular}{ccccccccc}
\hline 
\hline Mode & $\left(1,1\right)$ & $\left(1,2\right)$ & $\left(1,3\right)$ & $\left(1,4\right)$ & $\left(2,1\right)$ & $\left(2,2\right)$ & $\left(2,3\right)$ & $\left(3,3\right)$\tabularnewline
\hline
$\bar{v}_{{\rm circ}}$ $[{\rm km}/{\rm s}]$ & 1032 & 1296 & 1569 & 1778 & 1364 & 1661 & 2002 & 2444\tabularnewline
\hline
\end{tabular}
\end{table}

\subsection{Constant $f\left(\psi\right)$}

\begin{figure}
\noindent \centering{}\hspace{-.5cm}\includegraphics[scale=0.3]{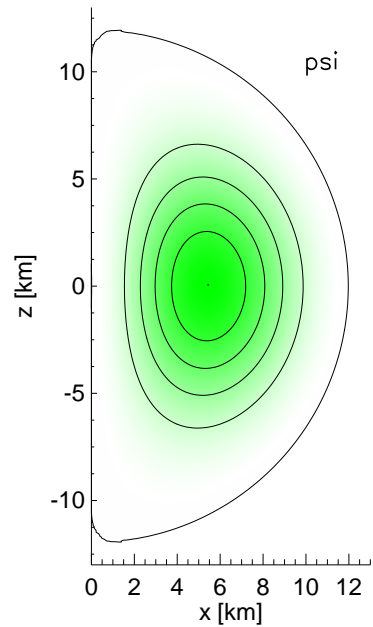}\caption{\textbf{Constant $f\left(\psi\right)$}. Stream function $\psi$ for
the case $n=0$. The color coding is the same one as in Fig. \ref{fig:Initial}.\label{fig:Constant}}

\end{figure}

Eriguchi \textit{et al.~}\cite{1986.Ewald} investigated both cases
(\ref{eq:77}). Therefore, we now take the reference model and set
the parameter $n=0$ such that we obtain\[
f\left(\psi\right)=k.\]
For this case, we do not find a collection of circulation modes but
only a single solution, which is shown in Fig. \ref{fig:Constant}.
There, we see that compared to the upper left panel of Fig. \ref{fig:Modes-1}
the stream function is slightly deformed. In addition to that, we
find that the average circulation velocity becomes\[
\bar{v}_{{\rm circ}}=1366\,{\rm km}/{\rm s},\]
while the average proper radius $\bar{R}$ and the mass ${\cal M}$
of the neutron star are the same as for the reference model.

\subsection{Rotating neutron star}

In contrast to Eriguchi \textit{et al.~}\cite{1986.Ewald}, we have
also computed rotating neutron star models with meridional circulation.
For that purpose, we set the specific angular momentum $L\left(\psi\right)$
to a non-vanishing value. The simplest choice would be\begin{equation}
L\left(\psi\right)={\rm const}\neq0,\label{eq:81}\end{equation}
where the neutron star rotates rapidly near the rotation axis. Eriguchi
\textit{et al.~}\cite{1986.Ewald} exclude such models, because the
rotation velocity becomes infinite when approaching the rotation axis
in Newtonian configurations. In general relativity, the situation
is more complicated, because spacetime itself can be dragged by the
neutron star fluid. We are not able to present stringent analytical
reasons which exclude condition (\ref{eq:81}), but simulations obeying
that condition do not seem to converge for higher numerical resolutions
and have to be considered as invalid neutron star models.

\begin{figure}
\noindent \centering{}\hspace{-3.8cm}\includegraphics[scale=0.3]{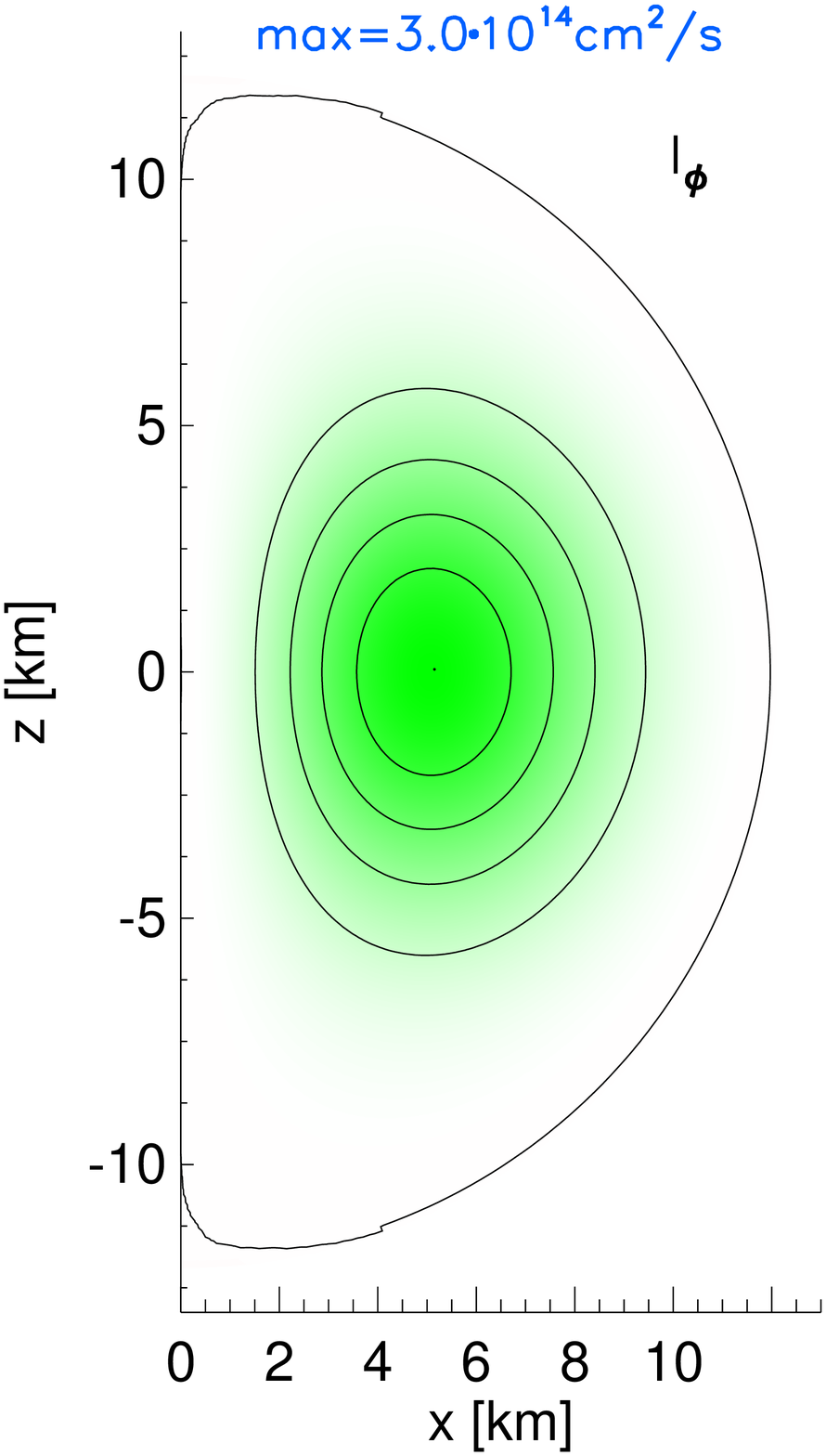}\put(-62.5,0){\includegraphics[scale=0.3]{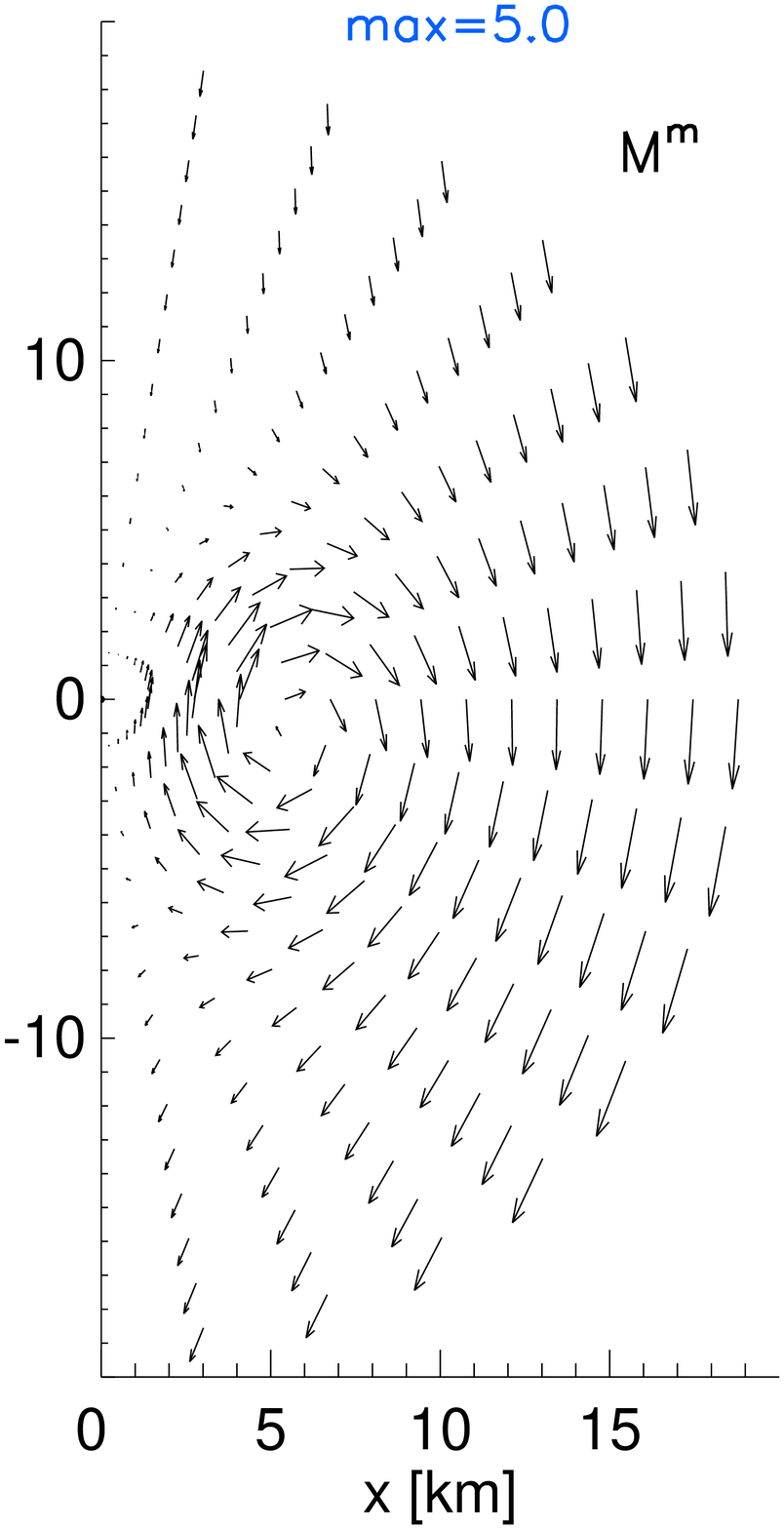}}\caption{\textbf{Rotating neutron star}. Specific angular momentum $l_{\phi}$
and the $2$-shift $M^{m}$ for the rotating model (\ref{eq:82}).
The color coding of the contour plot is the same one as in Fig. \ref{fig:Initial}.
For the vector plot, the quantity $max$ denotes the maximum vector
length.\label{fig:Rotating}}

\end{figure}

Therefore, we consider the next simplest case\[
L\left(\psi\right)={\rm const}\cdot\psi.\]
For the choice\begin{equation}
L\left(\psi\right)=3\cdot10^{14}\frac{\psi}{\psi_{{\rm max}}}\frac{{\rm cm}^{2}}{{\rm s}},\label{eq:82}\end{equation}
we find the average circulation and rotation velocities\begin{eqnarray}
\bar{v}_{{\rm circ}} & = & 1168\,{\rm km}/{\rm s},\nonumber \\
\bar{v}_{{\rm rot}} & = & 431\,{\rm km}/{\rm s},\label{eq:83}\end{eqnarray}
where the latter is the average value of\[
v_{{\rm rot}}=r\sin\theta v^{\phi}.\]
When compared to the reference model, the average radius and the mass
do not show significant changes, just as most basic fields. Two exceptions
are the specific angular momentum $l_{\phi}$ and the $2$-shift $M^{m}$,
which vanish for the reference model, whereas for the rotating model
(\ref{eq:82}) their distributions look like as displayed in Fig.
\ref{fig:Rotating}. They now exhibit a vortex, similar to the stream
function $\psi$ and the $3$-shift $N^{a}$. In contrast to the reference
model, the $3$-shift $N^{a}$ vector does not only lie within the
meridional plane as shown in the lower left panel of Fig. \ref{fig:Reference},
but it has also contributions perpendicular to that plane for the
rotating model (not displayed in this paper).

Having investigated model (\ref{eq:82}), where the circulation and
rotation velocities are of the same order of magnitude, the question
arises what happens when the rotational velocity is much smaller or
much larger than the circulation velocity. Since the reference model
itself is non-rotating, we continuously increased its rotation velocity
from zero to the value (\ref{eq:83}) by changing the constant in
relation (\ref{eq:82}). Currently, GRNS fails to compute models where
the average rotation velocity is larger than about the value (\ref{eq:83}).
Actually, one would expect that there are at least a few models where
the rotation velocity is much larger than the circulation velocity,
and possible improvements in the numerical method could allow their
computation.%

\subsection{Higher modes with circulation and rotation}

There are not only higher circulation modes for non-rotating configurations,
like the reference model, but also for rotating models. In this case,
the difference between the fundamental mode and the higher modes is
similar to what has been discussed in Sect. \ref{sub:Higher-circulation-modes}.
All of the basic fields containing a single vortex for the fundamental
mode, i.e. the four fields $\psi,N^{a},l_{\phi},M^{m}$, exhibit the
same number of vortices for a certain higher mode. These vortices
are always roughly located at the same spatial position for the four
fields. Similar to the non-rotating configurations, the modes where
both circulation and rotation are present turn out to be a $2$-dimensional
mode collection.

\begin{figure}
\noindent \centering{}\hspace{-.3cm}\includegraphics[scale=0.25]{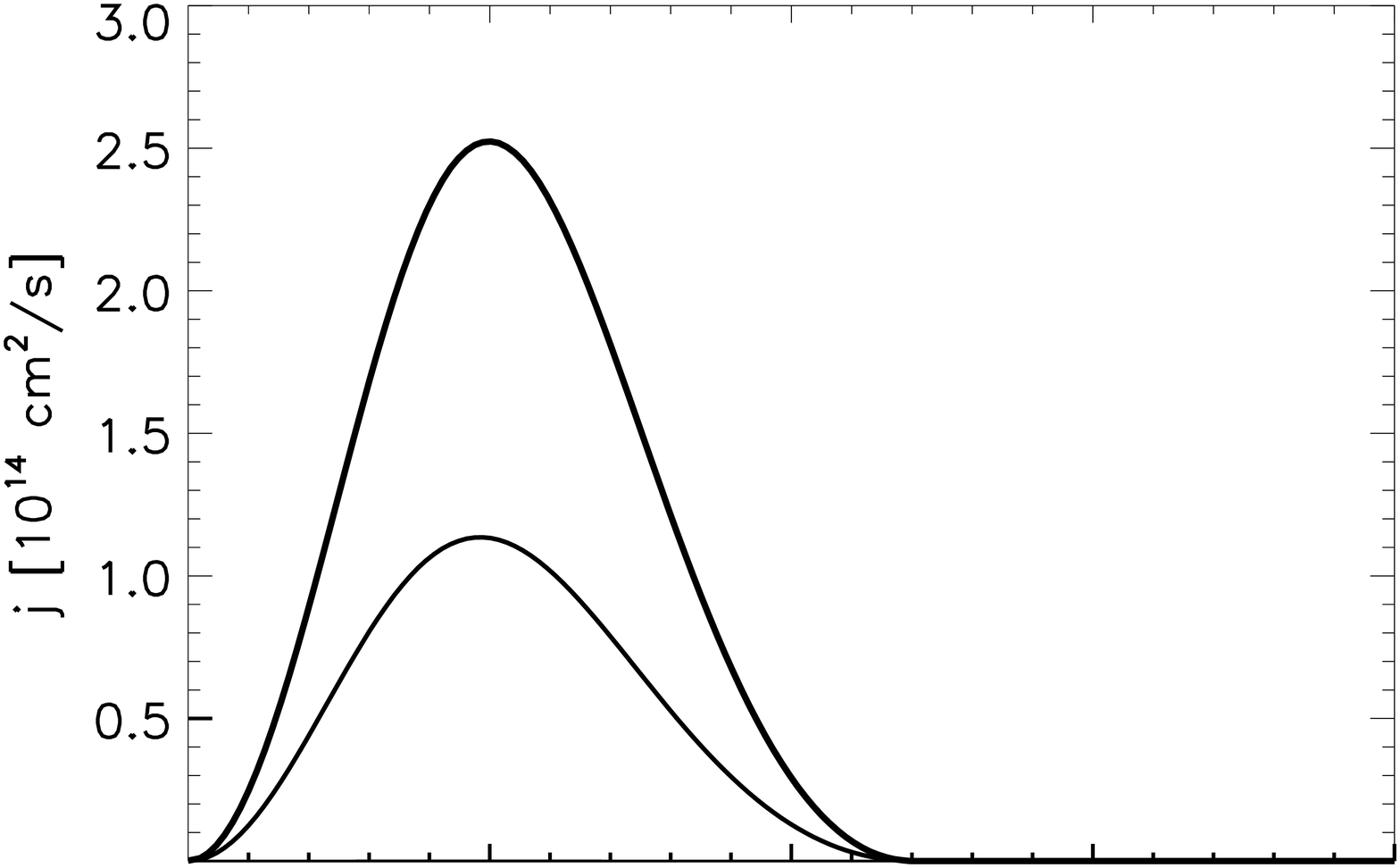}\\
\vspace{-1.55cm}\hspace{-.3cm}\includegraphics[scale=0.25]{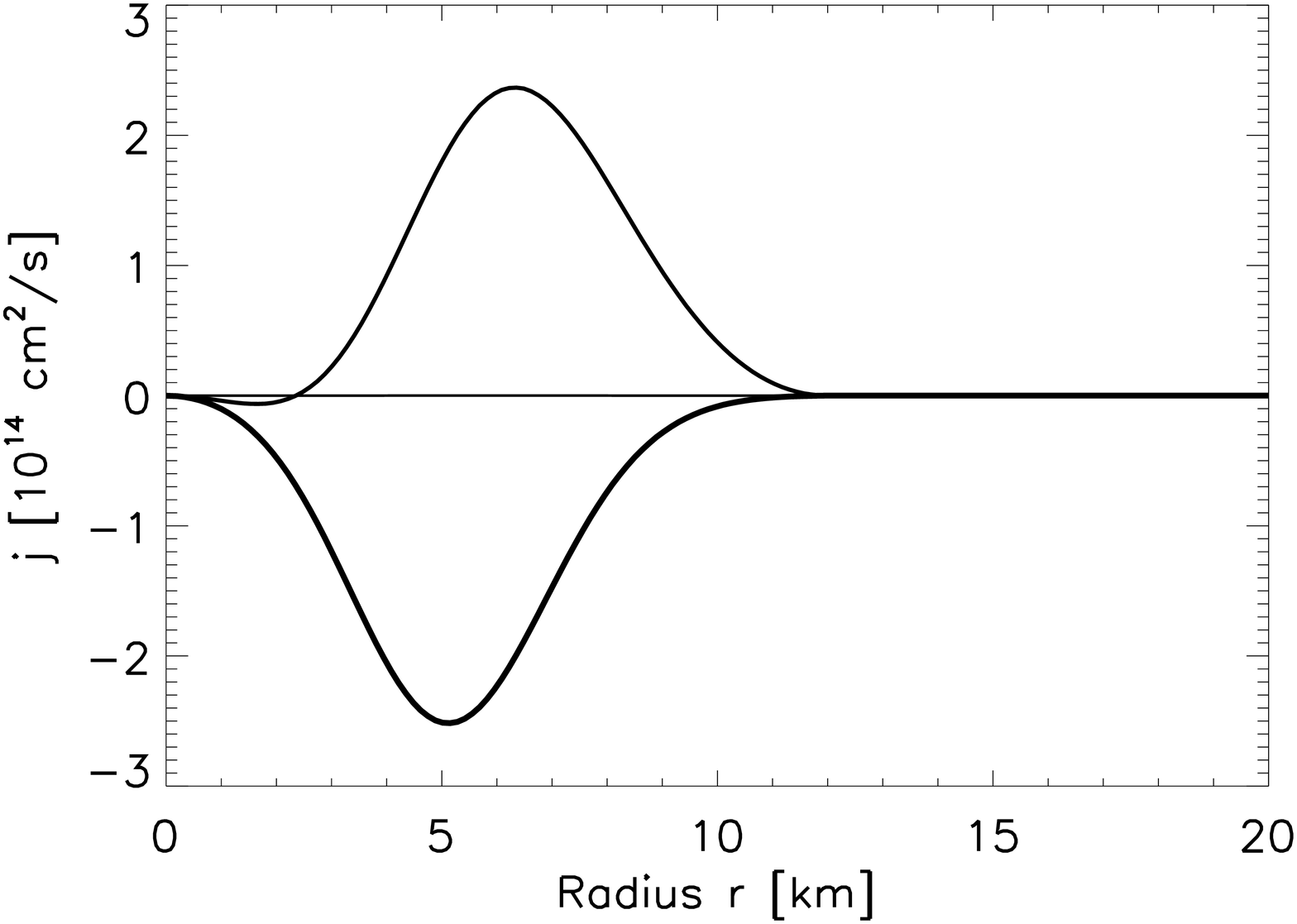}\caption{\textbf{Specific angular momentum.} The two panels show radial profiles
of the specific angular momentum $j$ for two different modes of the
rotating model (\ref{eq:82}) at angles $\theta=0$ (thinnest curve),
$\theta=\frac{\pi}{4}$, and $\theta=\frac{\pi}{2}$ (thickest curve),
respectively. The upper panel belongs to the fundamental mode, displayed
in Fig. \ref{fig:Rotating}, and the lower one to one of the higher
modes. The specific angular momentum $l_{\phi}$ and the stream function
$\psi$ of this higher mode exhibit a vortex distribution as in the
upper, left panel of Fig. \ref{fig:Modes-2}.\label{fig:Profile}}

\end{figure}

Fig. \ref{fig:Profile} shows radial profiles of the specific angular
momentum (per unit rest mass)\[
j=\frac{\epsilon+p}{\rho}u_{\phi}\]
(different from the specific angular momentum $l_{\phi}$, which is
defined per unit energy) for the fundamental and one higher mode belonging
to the rotating model (\ref{eq:82}). There, we do not only recognize
combined co- and counterrotation, but also that in parts of the star
${\rm dj}/{\rm dr}<0$, which implies that the flow does not satisfy
Solberg's criterion for local stability (see \cite{1975.Seguin} for
a proof in GR). %

\section{Conclusions\label{sec:Conclusions}}

We computed the first stationary, axisymmetric neutron star models
with meridional circulation in the framework of general relativity.
For that purpose, we have developed GRNS, a new code that uses a fixed
point iteration method starting from a Tolman-Oppenheimer-Volkoff
like initial configuration, similar to RNS, written by N. Stergioulas.

We found meridional circulation modes like Eriguchi \textit{et al.~}\cite{1986.Ewald}
in the Newtonian framework. However, by using a fixed point iteration
instead of a Newton-Raphson one, we were able to automatize the process
of computing such modes. As a result, we identified a much larger
set of modes than these authors. Our study shows that the circulation
models form a two-dimensional set.

The circulation modes differ by a varying number of vortices in the
stream lines of the fluid. These vortices cause deformations of the
surface of the neutron star. The deformations are influenced strongest
by the vortices in the vicinity of the surface, and their influence
rises with the value of the circulation velocity. We found surface
deformations of the order of one percent for expected maximal circulation
velocities of about $1000\mbox{\,}{\rm km}/{\rm s}$. However, the
radius and mass of the neutron star do not significantly depend on
the circulation velocities.

In contrast to Eriguchi \textit{et al.} \cite{1986.Ewald}, we also
computed rotating neutron star models with meridional circulation.
For such models, the rotation velocity is highest in the center of
the vortices, and vanishes at the surface of the neutron star. In
addition to that, we were unable to find rotating neutron star models,
where the rotation velocity is significantly larger than the circulation
velocity. We are not sure whether this is possibly caused by our symmetry
assumptions.

There are clear perspectives for a future application of the outcomes
of this investigation. Perturbing the obtained modes, a dynamical
evolution of the neutron star could show the influence of meridional
circulations on gravitational waves, for which a direct experimental
detection is expected in the near future. Another application is investigating
the influence of meridional circulations on neutron star oscillations.
Both methods offer a way to determine by means of observations whether
meridional circulations are present in (young) neutron stars. At the
current stage we are unable to decide how common such circulations
are in nature, because with our study we did not evaluate stability
criteria for the circulation modes.

In the near future, the two most important extensions of this investigation
will be changing the topology to a toroidal one and to include magnetic
fields. We have already presented some first thoughts in that direction
in the Appendix finding that the field equation for the specific angular
momentum is strongly affected by the presence of a magnetic field.

\section*{ACKNOWLEDGMENTS}

We are grateful to Eric Gourgoulhon for useful comments. This work
was supported by the DAAD via an IKYDA German--Greek research travel
grant, by the Collaborative Research Center on Gravitational Wave
Astronomy of the Deutsche Forschungsgemeinschaft (DFG SFB/Transregio
7), and by CompStar, a Research Networking Programme of the European
Science Foundation.

\appendix

\section{Corrected geometry equations\label{sec:Corrected-geometry-equations}}

In our investigation, we use equations (B3-B7) of Gourgoulhon \textit{et
al.~}\cite{1993.Gourgoulhon}, which contain minor errors in the
three equations (B3), (B4a), and (B4b). We found these errors by thoroughly
verifying all the mathematical equations in that paper. For equation
(B3), this was done by hand, and similar to Gourgoulhon \textit{et
al.~}\cite{1993.Gourgoulhon} we used the computer algebra program
Mathematica to validate the rather lengthy equations (B4a) and (B4b).
In the following, we will give the three corrected equations. Due
to their length, we will not list them completely here, but show only
where the corrections appear. Several new mathematical quantities
are introduced during the derivation of the equations, which will
not be defined here. Instead, we refer to Gourgoulhon \textit{et al.~}\cite{1993.Gourgoulhon},
whose notation and conventions are adopted by us, except for the differently
chosen indices (\ref{eq:1}).

Equation (B3) of Gourgoulhon \textit{et al.~}\cite{1993.Gourgoulhon}
has the form\[
...=4\pi\left(E+S_{a}^{a}\right)+K_{ab}K^{ab}+\frac{L^{2}}{2},\]
 where the term $2m^{r}m^{\theta}\nu_{,r}\nu_{,\theta}$ is missing
on the left hand side, i.e. the correct equation (B3) reads\[
...+2m^{r}m^{\theta}\nu_{,r}\nu_{,\theta}=4\pi\left(E+S_{a}^{a}\right)+K_{ab}K^{ab}+\frac{L^{2}}{2}.\]
In the last line of the left hand side of equation (B4a) of Gourgoulhon
\textit{et al.~}\cite{1993.Gourgoulhon}, there appears the expression\[
...+\frac{m^{\theta}}{M}\left[M_{\;\;,\theta}^{r}\left(\mu_{,\theta}-4\alpha_{,\theta}\right)-\frac{M_{\;\;,\theta\theta}^{r}}{M}\right]...\,.\]
Here, the presence of the $2$-lapse $M$ in the squared bracket is
an error, i.e. the correct equation (B4a) has the form\[
...+\frac{m^{\theta}}{M}\left[M_{\;\;,\theta}^{r}\left(\mu_{,\theta}-4\alpha_{,\theta}\right)-M_{\;\;,\theta\theta}^{r}\right]...\,.\]
The second but last line of the left hand side of equation (B4b) of
Gourgoulhon \textit{et al.~}\cite{1993.Gourgoulhon} reads\[
...\frac{m^{\theta}}{M}\left[...-\frac{M_{\;\;,r\theta}^{r}}{M}-2\frac{M_{\;\;,\theta\theta}^{\theta}}{M}-2\frac{A_{,\theta\theta}}{A}\right]...\,.\]
The presence of the $2$-lapse $M$ in the squared bracket is an error,
and the third term $-2A_{,\theta\theta}/A$ has to be deleted such
that the correct equation (B4b) reads\[
...\frac{m^{\theta}}{M}\left[...-M_{\;\;,r\theta}^{r}-2M_{\;\;,\theta\theta}^{\theta}\right]...\,.\]

\section{Flat-space Laplacians\label{sec:Flat-space-Laplacians}}

In the following, we consider $2$- and $3$-dimensional Laplacians.
For that purpose, we use the inverse $2$- and $3$-metrices $k^{mn}$
and $h^{ab}$ \cite{1993.Gourgoulhon}, and the Christoffel symbols
of the second\begin{eqnarray*}
^{2}\Gamma_{no}^{m} & = & k^{mp}\,{}\Gamma_{pno},\\
^{3}\Gamma_{bc}^{a} & = & h^{ad}\,{}\Gamma_{dbc}\end{eqnarray*}
and of the first kind\[
\Gamma_{\alpha\beta\gamma}=\frac{1}{2}\left(\partial_{\beta}g_{\alpha\gamma}+\partial_{\gamma}g_{\beta\alpha}-\partial_{\alpha}g_{\beta\gamma}\right).\]
These Christoffel symbols allow us to compute the $2$- and $3$-covariant
derivatives \cite{1993.Gourgoulhon} of a tensor $T_{b...}^{a...}$:\begin{eqnarray*}
T_{n...||p}^{m...} & = & \partial_{p}T_{n...}^{m...}+^{2}\Gamma_{qp}^{n}T_{m...}^{q...}+...-{}^{2}\Gamma_{np}^{q}T_{q...}^{m...}-...,\\
T_{b...|c}^{a...} & = & \partial_{c}T_{b...}^{a...}+^{3}\Gamma_{dc}^{a}T_{b...}^{d...}+...-{}^{3}\Gamma_{bc}^{d}T_{d...}^{a...}-...\,.\end{eqnarray*}
Then, limiting ourselves to a scalar $\Phi$ and a vector $\Phi^{a}$,
we construct the following Laplacians:\begin{eqnarray*}
k^{np}\Phi_{||np} & = & ^{2}\Delta\Phi,\\
h^{bc}\Phi_{|bc} & = & ^{3}\Delta\Phi,\\
k^{np}\Phi_{\;\;\;||np}^{m} & = & ^{2}\Delta_{\;\;\; n}^{m}\Phi^{n},\\
h^{bc}\Phi_{\;\;|bc}^{a} & = & ^{3}\Delta_{\;\; b}^{a}\Phi^{b}.\end{eqnarray*}
For flat-space, we choose\begin{eqnarray*}
k_{mn} & = & {\rm diag}\left(1,r^{2}\right),\\
h_{ab} & = & {\rm diag}\left(1,r^{2},r^{2}\sin^{2}\theta\right),\end{eqnarray*}
such that a straightforward calculation gives the two well known scalar
Laplacians\begin{eqnarray*}
^{2}\Delta & = & \partial_{r}^{2}+\frac{1}{r^{2}}\partial_{\theta}^{2}+\frac{1}{r}\partial_{r},\\
^{3}\Delta & = & \partial_{r}^{2}+\frac{1}{r^{2}}\partial_{\theta}^{2}+\frac{2}{r}\partial_{r}+\frac{\cot\theta}{r^{2}}\partial_{\theta}+\frac{1}{r^{2}\sin^{2}\theta}\partial_{\phi}^{2},\end{eqnarray*}
and the two less familiar vector Laplacians\begin{eqnarray}
^{2}\Delta_{\;\;\; n}^{m} & = & \delta_{n}^{m}\,{}^{2}\Delta+\left(\begin{array}{cc}
-\frac{1}{r^{2}} & -\frac{2}{r}\partial_{\theta}\\
\frac{2}{r^{3}}\partial_{\theta} & \frac{2}{r}\partial_{r}\end{array}\right),\nonumber \\
^{3}\Delta_{\;\; b}^{a} & = & \delta_{b}^{a}\,{}^{3}\Delta+\left(\begin{array}{ccc}
-\frac{2}{r^{2}} & -\frac{2}{r}\partial_{\theta}-2\frac{\cot\theta}{r} & 0\\
\frac{2}{r^{3}}\partial_{\theta} & \frac{2}{r}\partial_{r}+\frac{1-\cot^{2}\theta}{r^{2}} & 0\\
0 & 0 & m_{33}\end{array}\right),\;\;\;\;\;\;\;\;\label{eq:36}\end{eqnarray}
with the matrix element $m_{33}=\frac{2}{r}\partial_{r}+2\frac{\cot\theta}{r^{2}}\partial_{\theta}$.
Note that equation (\ref{eq:36}) holds only in case of axisymmetry,
where the azimuthal derivatives $\partial_{\phi}$ of the Laplacians
$^{3}\Delta_{\;\; b}^{a}$ and $^{3}\Delta$ vanish.

\section{Energy equation\label{sec:Energy-equation}}

In this appendix, we show that for our symmetry assumptions it is
possible to write the energy equation (\ref{eq:10}) as the vanishing
flat-space $3$-divergence (\ref{eq:19}) of the $3$-vector (\ref{eq:25}).
For that purpose, we expand equation (\ref{eq:10}) as\begin{equation}
\partial_{\alpha}\left[\left(\epsilon+p\right)u^{\alpha}\right]+\left(\epsilon+p\right)\Gamma_{\beta\alpha}^{\beta}u^{\alpha}=u^{\alpha}\partial_{\alpha}p,\label{eq:13}\end{equation}
where $\Gamma_{\beta\gamma}^{\alpha}=g^{\alpha\delta}\Gamma_{\delta\beta\gamma}$
are the common $4$-dimensional Christoffel symbols of the second
kind. Because of~\cite{1992.DInverno}\[
\Gamma_{\beta\alpha}^{\beta}=\partial_{\alpha}\ln\sqrt{-g},\]
with the determinant $g=\det g_{\alpha\beta}$, and our assumptions
of stationarity and axisymmetry all temporal and azimuthal derivatives
$\partial_{t}...$ and $\partial_{\phi}...$ vanish. Hence, equation
(\ref{eq:13}) becomes\begin{equation}
\partial_{m}\left[\left(\epsilon+p\right)u^{m}\right]+\left(\epsilon+p\right)u^{m}\partial_{m}\ln\sqrt{-g}=u^{m}\partial_{m}p.\label{eq:18}\end{equation}
To compute the determinant $g$, we use its relation to the determinant
$k={\rm det}k_{mn}$ (see equation (2.27) of Gourgoulhon \textit{et
al.~}\cite{1993.Gourgoulhon}), the fact that in MTCMA coordinates
(Sect. \ref{sub:Basic-fields}) $k=A^{4}r^{2}$, and the definitions
(\ref{eq:14}) and (\ref{eq:16}):\[
\sqrt{-g}=NM\sqrt{k}=e^{2\alpha+\gamma}r^{2}\sin\theta.\]
For the right hand side of equation (\ref{eq:18}), we apply the temporal
component\begin{equation}
u^{m}\partial_{m}p=-\left(\epsilon+p\right)u^{m}\partial_{m}\ln u_{t}\label{eq:26}\end{equation}
of the general relativistic Euler equation (\ref{eq:23}), obtained
by setting $\alpha=t$ in that equation and using stationarity. Hence,
equation (\ref{eq:18}) becomes\begin{eqnarray*}
 &  & \;\;\,\partial_{m}\left[e^{2\alpha+\gamma}\left(\epsilon+p\right)u_{t}u^{m}\right]\\
 &  & +e^{2\alpha+\gamma}\left(\epsilon+p\right)u_{t}\left(\frac{2}{r}u^{r}+\cot\theta u^{\theta}\right)=0\end{eqnarray*}
which is equal to equation (\ref{eq:19}).

\section{Meridional components\label{sec:Meridional-components}}

Let us consider the general relativistic Euler equation (\ref{eq:11})
rewritten in the form of equation (\ref{eq:27}). This appendix deals
with the meridional components of that equation, which means setting
$a=m$ in equation (\ref{eq:27}) such that\begin{equation}
\frac{\partial_{m}p}{\epsilon+p}=\frac{1}{2}u^{\alpha}u^{\beta}\partial_{m}g_{\alpha\beta}+u_{t}u^{n}\partial_{n}l_{m}.\label{eq:29}\end{equation}
For the second term on the right hand side, we write\[
u^{n}\partial_{n}l_{m}=rw\left(\begin{array}{c}
u^{\theta}\\
-u^{r}\end{array}\right)+u^{n}\partial_{m}l_{n},\]
with\begin{equation}
w=\frac{1}{r}\left(\partial_{\theta}l_{r}-\partial_{r}l_{\theta}\right),\label{eq:33}\end{equation}
begin the quantity `$\omega$' defined in Eriguchi \textit{et al.~}\cite{1986.Ewald}
in the Newtonian limit. Then, using the constraint (\ref{eq:22})
a short calculation shows that\[
\frac{1}{2}u^{\alpha}u^{\beta}\partial_{m}g_{\alpha\beta}-u_{t}u^{n}\partial_{m}\frac{u_{n}}{u_{t}}=-\partial_{m}\ln u_{t}+u_{t}u^{\phi}\partial_{m}\frac{u_{\phi}}{u_{t}}.\]
Hence, equation (\ref{eq:29}) can be brought in the form (\ref{eq:30}),
which has a form similar to equations (7) and (8) of Eriguchi \textit{et
al.~}\cite{1986.Ewald}.

\section{Special Laplacian\label{sec:Asymmetric-Laplacian}}

In the following, we show that there is a Laplacian hidden behind
the quantity $w$ defined in equation (\ref{eq:33}). For that purpose,
we extend equations (15) and (17) of Eriguchi \textit{et al.~}\cite{1986.Ewald}
to general relativity.

The first step is to define the quantity\[
D=\partial_{r}^{2}\psi+\frac{\sin\theta}{r^{2}}\partial_{\theta}\left(\frac{1}{\sin\theta}\partial_{\theta}\psi\right),\]
which is equal to the left hand side of equation (15) of Eriguchi
\textit{et al.~}\cite{1986.Ewald}. Then, equation (\ref{eq:20})
allows us to write\begin{equation}
D=\sin\theta\left[-\partial_{r}\left(r^{2}\varrho u^{\theta}\right)+\partial_{\theta}\left(\varrho u^{r}\right)\right].\label{eq:37}\end{equation}
Next, we use $u^{m}=g^{m\alpha}u_{\alpha}$, and the decompositions
(2.9b) and (2.25b) of Gourgoulhon \textit{et al.~}\cite{1993.Gourgoulhon}
to obtain after a short computation\begin{equation}
u^{m}=u_{t}k^{mn}\left(\frac{c_{n}}{\tilde{\varrho}}-l_{n}\right),\label{eq:38}\end{equation}
with\begin{eqnarray*}
c_{m} & = & \frac{\tilde{\varrho}}{u_{t}}k_{mn}\left[\frac{M^{n}}{M^{2}}\left(M^{p}u_{p}+u_{\phi}\right)-\frac{N^{n}N^{a}}{N^{2}}u_{a}-\frac{N^{n}}{N^{2}}u_{t}\right],\\
\tilde{\varrho} & = & \frac{\varrho u_{t}}{A^{2}}.\end{eqnarray*}
The only non-vanishing components of the $2$-metric $k^{mn}$ are
$k^{rr}=1/A^{2}$ and $k^{\theta\theta}=1/\left(A^{2}r^{2}\right)$.
Therefore, equation (\ref{eq:38}) gives\begin{eqnarray}
\partial_{r}\left(r^{2}\varrho u^{\theta}\right) & = & \partial_{r}c_{\theta}-\partial_{r}\left(\tilde{\varrho}l_{\theta}\right),\label{eq:39}\\
\partial_{\theta}\left(\varrho u^{r}\right) & = & \partial_{\theta}c_{r}-\partial_{\theta}\left(\tilde{\varrho}l_{r}\right).\label{eq:40}\end{eqnarray}
By inverting equation (\ref{eq:38}) to \[
l_{m}=\frac{1}{u_{t}}\left(\frac{u_{t}}{\tilde{\varrho}}c_{m}-k_{mn}u^{n}\right)\]
and using equation (\ref{eq:20}) we also find\begin{eqnarray}
l_{r} & = & \frac{1}{\tilde{\varrho}}\left(c_{r}-\frac{\partial_{\theta}\psi}{r^{2}\sin\theta}\right),\label{eq:41}\\
l_{\theta} & = & \frac{1}{\tilde{\varrho}}\left(c_{\theta}+\frac{\partial_{r}\psi}{\sin\theta}\right).\label{eq:42}\end{eqnarray}
Now we combine equations (\ref{eq:33}) and (\ref{eq:39}-\ref{eq:42})
such that equation (\ref{eq:37}) becomes\begin{eqnarray}
D & = & \sin\theta\left(\partial_{\theta}c_{r}-\partial_{r}c_{\theta}\right)-\tilde{\varrho}r\sin\theta w\label{eq:43}\\
 &  & +\left[\left(\partial_{r}\psi+c_{\theta}\sin\theta\right)\partial_{r}+\left(\frac{\partial_{\theta}\psi}{r^{2}}-c_{r}\sin\theta\right)\partial_{\theta}\right]\ln\tilde{\varrho},\nonumber \end{eqnarray}
which extends equation (15) of Eriguchi \textit{et al.~}\cite{1986.Ewald}
to general relativity.

Next, we introduce the quantity \cite{1986.Ewald}\[
\chi=\frac{\psi\cos\phi}{r\sin\theta}\]
to write equation (\ref{eq:43}) in the form\begin{eqnarray*}
D & = & \frac{r\sin\theta}{\cos\phi}\left\{ \frac{\cos\phi}{r}\left(\partial_{\theta}c_{r}-\partial_{r}c_{\theta}\right)-\tilde{\varrho}w\cos\phi\right.\\
 &  & \left.+\left[\left(\partial_{r}\chi+\frac{\chi}{r}+\frac{c_{\theta}}{r}\cos\phi\right)\partial_{r}\right.\right.\\
 &  & \left.\left.+\frac{1}{r^{2}}\left(\partial_{\theta}\chi+\chi\cot\theta-rc_{r}\cos\phi\right)\partial_{\theta}\right]\ln\tilde{\varrho}\right\} .\end{eqnarray*}
In the Newtonian limit, the expression in curly brackets becomes the
right hand side of equation (17) of Eriguchi \textit{et al.~}\cite{1986.Ewald},
i.e. equation (\ref{eq:35}) extends equation (17) of Eriguchi \textit{et
al.~}\cite{1986.Ewald} to general relativity under condition (\ref{eq:34}).

\section{Integral expansion\label{sec:Integral-expansion}}

Below, we expand the four integrals appearing in Sect. \ref{sub:Green-functions}
in terms of trigonometric functions and Legendre polynomials.

\subsection{2-scalar\label{sub:2-scalar}}

For the integral (\ref{eq:62}) we use the expansion\begin{eqnarray*}
\ln\left|\vec{x}-\vec{x}'\right| & = & \ln\max\left(r,r'\right)-\sum_{l=1}^{\infty}\frac{1}{l}\frac{\min^{l}\left(r,r'\right)}{\max^{l}\left(r,r'\right)}\\
 &  & \cdot\left(\cos\left(l\theta\right)\cos\left(l\theta'\right)+\sin\left(l\theta\right)\sin\left(l\theta'\right)\right)\end{eqnarray*}
of Komatsu \textit{et al.~}\cite{1989.Komatsu}. Applying the von
Neumann boundary condition\[
S\left(r,\pi+\theta\right)=S\left(r,\pi-\theta\right),\]
a short computation shows that\begin{eqnarray}
\Phi\left(r,\theta\right) & = & \frac{1}{\pi}\int_{0}^{\infty}{\rm d}r'r'\ln\max\left(r,r'\right)\int_{0}^{\pi}{\rm d}\theta'S\left(r',\theta'\right)\nonumber \\
 &  & -\frac{1}{\pi}\sum_{l=1}^{\infty}\frac{1}{l}\cos\left(l\theta\right)\int_{0}^{\infty}{\rm d}r'r'\frac{\min^{l}\left(r,r'\right)}{\max^{l}\left(r,r'\right)}\;\;\;\;\;\;\;\;\;\;\label{eq:71}\\
 &  & \;\;\;\;\;\;\;\;\;\;\;\;\;\;\;\;\;\;\;\;\;\;\;\;\;\;\,\cdot\int_{0}^{\pi}{\rm d}\theta'\cos\left(l\theta'\right)S\left(r',\theta'\right).\nonumber \end{eqnarray}
We use this expansion for the basic geometry field $\alpha$, i.e.
equation (\ref{eq:53}). In that case, the potential is $\Phi=\alpha+\nu$
and the source $S=S_{\alpha}$.

For the basic geometry field $\beta$, i.e. equation (\ref{eq:54}),
the integral (\ref{eq:62}) has to be evaluated. The potential is
$\Phi=r\sin\theta\left(\beta+\nu\right)$, which has to vanish on
the rotation axis. Therefore, we apply the Dirichlet boundary condition\[
S\left(r,\pi+\theta\right)=-S\left(r,\pi-\theta\right)\]
and find\begin{eqnarray}
\Phi\left(r,\theta\right) & = & -\frac{1}{\pi}\sum_{l=1}^{\infty}\frac{1}{l}\sin\left(l\theta\right)\int_{0}^{\infty}{\rm d}r'r'\frac{\min^{l}\left(r,r'\right)}{\max^{l}\left(r,r'\right)}\;\;\;\;\;\;\;\;\;\;\label{eq:70}\\
 &  & \;\;\;\;\;\;\;\;\;\;\;\;\;\;\;\;\;\;\;\;\;\;\;\;\;\;\,\cdot\int_{0}^{\pi}{\rm d}\theta'\sin\left(l\theta'\right)S\left(r',\theta'\right),\nonumber \end{eqnarray}
with the source $S=S_{\beta}$.

\subsection{3-scalar}

For the integral (\ref{eq:63}) we use the expansion \cite{1989.Komatsu}\begin{eqnarray*}
\frac{1}{\left|\vec{x}-\vec{x}'\right|} & = & \sum_{l=0}^{\infty}\sum_{m=-l}^{l}\frac{\left(l-m\right)!}{\left(l+m\right)!}\frac{\min^{l}\left(r,r'\right)}{\max^{l+1}\left(r,r'\right)}\\
 &  & \;\;\;\;\;\;\;\;\;\;\;\;\;\;\;\cdot P_{l}^{m}\left(\cos\theta'\right)P_{l}^{m}\left(\cos\theta\right)e^{im\left(\phi-\phi'\right)},\end{eqnarray*}
where $P_{l}^{m}$ are the associated Legendre polynomials. In case
of an axisymmetric source\[
S\left(\vec{x}\right)=S\left(r,\theta\right),\]
it is easy to show that\begin{eqnarray*}
\Phi\left(r,\theta\right) & = & -\frac{1}{2}\sum_{l=0}^{\infty}P_{l}\left(\cos\theta\right)\int_{0}^{\infty}{\rm d}r'r'^{2}\frac{\min^{l}\left(r,r'\right)}{\max^{l+1}\left(r,r'\right)}\\
 &  & \;\;\;\;\;\;\;\;\;\;\;\;\;\cdot\int_{0}^{\pi}{\rm d}\theta'P_{l}\left(\cos\theta'\right)\sin\theta'S\left(r',\theta'\right).\end{eqnarray*}
We apply this result to equation (\ref{eq:52}), and choose the potential
$\Phi=\nu$ together with the source $S=S_{\nu}$.

The integral (\ref{eq:63}) also appears in the equation for the basic
matter field $\chi_{0}$ (\ref{eq:57}). In that case, the source
is no longer axisymmetric but obeys\[
S\left(\vec{x}\right)=S\left(r,\theta\right)\cos\phi.\]
Then, it is possible to show that\begin{eqnarray}
\Phi\left(r,\theta,\phi\right) & = & \Phi\left(r,\theta\right)\cos\phi,\nonumber \\
\Phi\left(r,\theta\right) & = & \Phi_{0}\left(r,\theta\right)+...,\label{eq:64}\end{eqnarray}
with\begin{eqnarray}
\Phi_{0}\left(r,\theta\right) & = & -\frac{1}{2}\sum_{l=1}^{\infty}\frac{1}{l(l+1)}P_{l}^{1}\left(\cos\theta\right)\label{eq:67}\\
 &  & \;\;\;\;\;\;\;\;\;\;\;\;\,\cdot\int_{0}^{\infty}{\rm d}r'r'^{2}\frac{\min^{l}\left(r,r'\right)}{\max^{l+1}\left(r,r'\right)}\nonumber \\
 &  & \;\;\;\;\;\;\;\;\;\;\;\;\,\cdot\int_{0}^{\pi}{\rm d}\theta'P_{l}^{1}\left(\cos\theta'\right)\sin\theta'S\left(r',\theta'\right).\nonumber \end{eqnarray}
The dots appearing in equation (\ref{eq:64}) serve the same purpose
as in equation (\ref{eq:63}), i.e. they represent a remaining degree
of freedom for the choice of the boundary condition of the field $\chi_{0}$.
Looking at equation (19) of Eriguchi \textit{et al.~}\cite{1986.Ewald}
we actually see that\begin{equation}
\Phi\left(r,\theta\right)=\Phi_{0}\left(r,\theta\right)+\sum_{l=1}^{\infty}a_{l}r^{l}P_{l}^{1}\left(\cos\theta\right),\label{eq:66}\end{equation}
with arbitrary coefficients $a_{l}$. We have to choose these coefficients
in such a manner that there is no flow across the surface of the neutron
star. For that purpose, we consider surface-adapted coordinates $(\tilde{r},\tilde{\theta})$
(Sect. \ref{sub:Fixed-point-iteration}), where the radial coordinate
of the neutron star's surface becomes $\tilde{r}=1$ (\ref{eq:65}).
The condition for no flow across the surface is\[
\tilde{u}^{r}\stackrel{S}{=}0\]
(the letter `S' denotes that this relation holds only on the surface).
Using equation (\ref{eq:20}), it is straightforward to show that
this leads to\[
\partial_{\tilde{\theta}}\psi\stackrel{S}{=}0\]
As\[
\psi=r\sin\theta\chi_{0}=\tilde{r}R\left(\tilde{\theta}\right)\sin\tilde{\theta}\chi_{0}\]
we then find\begin{equation}
\left(1+\frac{R'\left(\tilde{\theta}\right)}{R\left(\tilde{\theta}\right)}\tan\tilde{\theta}\right)\chi_{0}+\tan\tilde{\theta}\partial_{\tilde{\theta}}\chi_{0}\stackrel{S}{=}0\label{eq:86}\end{equation}
Though not being the only mathematical solution, we currently limit
ourselves to the case\begin{equation}
\chi_{0}\stackrel{S}{=}0\label{eq:87}\end{equation}
which is the Dirichlet boundary condition and which leads to\[
\partial_{\tilde{\theta}}\chi_{0}\stackrel{S}{=}0\]
because in surface-adapted coordinates the surface corresponds to
a constant radial coordinate $\tilde{r}$. This shows that the Dirichlet
boundary condition (\ref{eq:87}) implies the constraint (\ref{eq:86})
for no flow across the surface. In terms of the potential\[
\tilde{\Phi}\left(\tilde{r},\tilde{\theta}\right)=\Phi\left(r,\theta\right)\]
equation (\ref{eq:87}) implies\[
\tilde{\Phi}\left(1,\tilde{\theta}\right)=0.\]
For this boundary condition, it can be shown that\begin{eqnarray}
\tilde{\Phi}\left(\tilde{r},\tilde{\theta}\right) & = & \tilde{\Phi}_{0}\left(\tilde{r},\tilde{\theta}\right)-\frac{1}{2}\sum_{l=1}^{\infty}\frac{2l+1}{l\left(l+1\right)}P_{l}^{1}\left(\cos\tilde{\theta}\right)\tilde{r}^{l}\label{eq:68}\\
 &  & \;\;\;\;\;\;\;\;\;\;\;\;\;\;\;\;\cdot\int_{0}^{\pi}{\rm d}\tilde{\theta}'P_{l}^{1}\left(\cos\tilde{\theta}'\right)\sin\tilde{\theta}'\tilde{\Phi}_{0}\left(1,\tilde{\theta}'\right),\nonumber \end{eqnarray}
with\[
\tilde{\Phi}_{0}\left(\tilde{r},\tilde{\theta}\right)=\Phi_{0}\left(r,\theta\right).\]
Note that we use the expansions (\ref{eq:67}) and (\ref{eq:68})
both for the potential $\Phi\left(r,\theta\right)=\chi_{0}\left(r,\theta\right)$
and the source $S\left(r,\theta\right)=S{}_{\chi_{0}}\left(r,\theta\right)$.

\subsection{2-vector}

Next, we consider the third integral of Sect. \ref{sub:Green-functions}
to evaluate the potential $\Phi^{m}=e^{2\left(\alpha+\nu\right)}M^{m}$
and the source $S^{m}=S_{M}^{m}$. Similar to Sect. \ref{sub:2-scalar},
we have to specify boundary conditions. We use the $2$-dimensional
Cartesian coordinates introduced in equation (\ref{eq:69}), in which
the potential and the source have the components $\left(\Phi^{x},\Phi^{z}\right)$
and $\left(S^{x},S^{z}\right)$, respectively. We apply the Dirichlet
boundary condition on the $x$-component $\Phi^{x}$, and the von
Neumann one on the $z$-component $\Phi^{z}$. These two quantities
are then governed by equations resulting from replacing\begin{eqnarray*}
\Phi\left(r,\theta\right) & \rightarrow & \Phi^{x}\left(r,\theta\right),\\
S\left(r',\theta'\right) & \rightarrow & S^{x}\left(r',\theta'\right)\end{eqnarray*}
in equation (\ref{eq:70}) and\begin{eqnarray*}
\Phi\left(r,\theta\right) & \rightarrow & \Phi^{z}\left(r,\theta\right),\\
S\left(r',\theta'\right) & \rightarrow & S^{z}\left(r',\theta'\right)\end{eqnarray*}
in equation (\ref{eq:71}).

\subsection{3-vector}

The fourth integral of Sect. \ref{sub:Green-functions} is used to
compute the potential $\Phi^{a}=N^{a}$ and the source $S^{a}=S_{N}^{a}$.
In this case, the derivation of the expansion in terms of Legendre
polynomials is somewhat lengthy, but still straightforward such that
we give only the result:\begin{eqnarray*}
\Phi^{a}\left(r,\theta\right) & = & -\frac{1}{2}\left(\begin{array}{cc}
\sin\theta & 0\\
\frac{\cos\theta}{r} & 0\\
0 & \frac{1}{r\sin\theta}\end{array}\right)\sum_{l=1}^{\infty}\frac{1}{l\left(l+1\right)}P_{l}^{1}\left(\cos\theta\right)\\
 &  & \;\;\cdot\int_{0}^{\infty}{\rm d}r'r'^{2}\frac{\min^{l}\left(r,r'\right)}{\max^{l+1}\left(r,r'\right)}\\
 &  & \;\;\cdot\int_{0}^{\pi}{\rm d}\theta'P_{l}^{1}\left(\cos\theta'\right)\sin\theta'\\
 &  & \;\;\cdot\left(\begin{array}{c}
\sin\theta'S^{r}\left(r',\theta'\right)+r'\cos\theta'S^{\theta}\left(r',\theta'\right)\\
r'\sin\theta'S^{\phi}\left(r',\theta'\right)\end{array}\right)^{T}\\
 &  & -\frac{1}{2}\left(\begin{array}{c}
\cos\theta\\
-\frac{\sin\theta}{r}\\
0\end{array}\right)\sum_{l=0}^{\infty}P_{l}\left(\cos\theta\right)\\
 &  & \;\;\cdot\int_{0}^{\infty}{\rm d}r'r'^{2}\frac{\min^{l}\left(r,r'\right)}{\max^{l+1}\left(r,r'\right)}\\
 &  & \;\;\cdot\int_{0}^{\pi}{\rm d}\theta'P_{l}\left(\cos\theta'\right)\sin\theta'\\
 &  & \;\;\cdot\left(\cos\theta'S^{r}\left(r',\theta'\right)-r'\sin\theta'S^{\theta}\left(r',\theta'\right)\right).\end{eqnarray*}

\section{Tests\label{sec:Tests}}

\begin{figure}
\centering{}\includegraphics[scale=0.33]{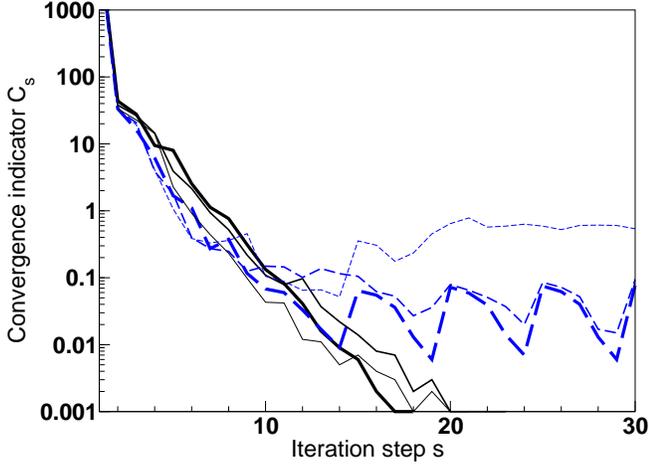}\caption{\textbf{Convergence tests}. Dependence of the convergence indicator
$C_{s}$ defined in equation (\ref{eq:79}) on the iteration step
$s$ for a model similar to the reference model. The three solid lines
refer to grid resolutions of $59\times52$, $150\times156$, and $501\times507$
zones (the higher the resolution, the thinner the corresponding line).
The blue dashed lines refer to the first higher mode.\label{fig:Resolution}}

\end{figure}
Here, we present convergence and consistency tests performed with
GRNS. For that purpose, we introduce the convergence indicator\begin{equation}
C_{s}=100\max_{F}\frac{\sum_{{\rm grid}}\left|F_{s}-F_{s-1}\right|}{\sum_{{\rm grid}}\left|F_{s}\right|},\label{eq:79}\end{equation}
where $F_{s}$ is the distribution of the basic field $F$ at the
iteration step $s$, and t The sums extend over the numerical grid.
First the fraction is evaluated for all basic fields $F$, and then
the maximum value determines the convergence indicator $C_{s}$. That
way, this quantity is most sensitive to the basic field which converges
least. Note that the convergence indicator becomes $C_{s}=0$ for
perfect convergence. Fig. \ref{fig:Resolution} shows convergence
tests for three different resolutions. We see that for the fundamental
mode GRNS converges for all three resolutions, i.e. the convergence
indicator $C_{s}$ approaches zero during the fixed point iteration.
For higher modes, the convergence indicator drops initially but then
starts to fluctuate, never getting close to the value zero. Improving
the weight function given in equation (\ref{eq:73}) might improve
this behavior. We have also considered different numbers of terms
$l_{{\rm max}}$ used in the sums (\ref{eq:80}), namely $3,10,50$,
which results is a convergence behavior similar to that shown in Fig.
\ref{fig:Resolution}, when the three grid resolutions are replaced
with the three values of $l_{{\rm max}}$. Eventually, we have validated
the correct behavior of GRNS, when the spatial extension of the numerical
grid is changed.

\section{Outlook on the magnetic field}

If a magnetic field is included, the stress-energy tensor (\ref{eq:84})
has the additional contribution\[
T_{\alpha\beta}^{{\rm EM}}=F_{\alpha\gamma}F_{\beta}^{\;\;\gamma}-\frac{1}{4}g_{\alpha\beta}F_{\gamma\delta}F^{\gamma\delta},\]
where $F_{\alpha\beta}=\partial_{\alpha}A_{\beta}-\partial_{\beta}A_{\alpha}$
is the electromagnetic field strength expressed in terms of the electromagnetic
$4$-vector potential $A_{\alpha}$. Moreover, we do not only have
to solve Einstein's field equation (\ref{eq:4}), but also Maxwell's
field equation\[
\nabla_{\alpha}F^{\alpha\beta}=\rho_{q}u^{\beta},\]
where $\rho_{q}$ is the charge density. A lengthy, but straightforward
computation shows that the general relativistic Euler equation (\ref{eq:11})
becomes\[
\left(\epsilon+p\right)u^{\beta}\nabla_{\beta}u^{\alpha}=-q^{\alpha\beta}\nabla_{\beta}p-\rho_{q}F_{\;\;\,\beta}^{\alpha}u^{\beta},\]
and equation (\ref{eq:28}) changes to\[
u^{m}\partial_{m}l_{\phi}=-\frac{\rho_{q}u^{m}\left(l_{\phi}\partial_{m}A_{t}+\partial_{m}A_{\phi}\right)}{u_{t}\left(\epsilon+p\right)}.\]
Hence, the simple solution method (\ref{eq:32}) does no longer work
if a magnetic field is present.\\

\section{Sources\label{sec:Sources}}

In Sect. \ref{sub:Geometry-equations}, we have written the covariant
Poisson equations (B3-B7) of Gourgoulhon \textit{et al.~}\cite{1993.Gourgoulhon}
to flat-space Poisson equations. The sources $S_{\nu}$, $S_{\alpha}$,
$S_{\beta}$, $S_{M}^{m}$, and $S_{N}^{a}$ of these equations, and
the source $S_{\chi_{0}}$ of the Poisson equation (\ref{eq:35})
are listed in the following. Similar to Appendix \ref{sec:Corrected-geometry-equations},
several new mathematical quantities appear below, which are defined
in Gourgoulhon \textit{et al.~}\cite{1993.Gourgoulhon}.

The scalar sources are\begin{widetext}\begin{eqnarray*}
S_{\nu} & = & A^{2}\left\{ 4\pi\left(E+S_{a}^{a}\right)+K_{ab}K^{ab}+\frac{L^{2}}{2}-\left[\frac{1}{A^{2}}+\left(m^{r}\right)^{2}\right]\left(\nu_{,r}\right)^{2}-\left[\frac{1}{\left(rA\right)^{2}}+\left(m^{\theta}\right)^{2}\right]\left(\nu_{,\theta}\right)^{2}-\left(m^{r}\right)^{2}\nu_{,rr}-\left(m^{\theta}\right)^{2}\nu_{,\theta\theta}\right.\\
 &  & \;\;\;\;\;\;\;\;\left.-2m^{r}m^{\theta}\nu_{,r\theta}-\left(m^{r}m_{\;\;,r}^{r}+m^{\theta}m_{\;\;,\theta}^{r}\right)\nu_{,r}-\left(m^{r}m_{\;\;,r}^{\theta}+m^{\theta}m_{\;\;,\theta}^{\theta}\right)\nu_{,\theta}-2m^{r}m^{\theta}\nu_{,r}\nu_{,\theta}\right\} -\beta_{,r}\nu_{,r}-\frac{\beta_{,\theta}\nu_{,\theta}}{r^{2}}\end{eqnarray*}
and\begin{eqnarray*}
S_{\alpha} & = & A^{2}\left\{ 8\pi s+\frac{1}{N}\left[\left(q^{r}+\omega m^{r}\right)\kappa_{,r}+\left(q^{\theta}+\omega m^{\theta}\right)\kappa_{,\theta}\right]+\frac{2}{MN}\left[\kappa_{r}[M,q]^{r}+\kappa_{\theta}[M,q]^{\theta}\right]+3\kappa_{m}\kappa^{m}\right.\\
 &  & \;\;\;\;\;\;\;\;\left.+\frac{1}{2}\left(\kappa_{mn}\kappa^{mn}+\kappa^{2}+L_{mn}L^{mn}\right)\right\} -\left(\nu_{,r}\right)^{2}-\left(\frac{\nu_{,\theta}}{r}\right)^{2}\end{eqnarray*}
and\begin{eqnarray*}
S_{\beta} & = & \frac{A^{2}}{e^{\gamma}}\left\{ 8\pi MNs_{m}^{m}-2\kappa_{r}\left[M,q\right]^{r}-2\kappa_{\theta}\left[M,q\right]^{\theta}-M\left(q^{r}+\omega m^{r}\right)\kappa_{,r}-M\left(q^{\theta}+\omega m^{\theta}\right)\kappa_{,\theta}\right.\\
 &  & \;\;\;\;\;\;\;\left.+MN\left(\kappa_{mn}\kappa^{mn}+\kappa^{2}-L_{mn}L^{mn}\right)\right\} -r\sin\theta\left[\left(\gamma_{r}\right)^{2}+\frac{1}{r^{2}}\left(\gamma_{,\theta}\right)^{2}\right]\end{eqnarray*}
and\[
S_{\chi_{0}}=\frac{1}{r}\left(\partial_{\theta}c_{r}-\partial_{r}c_{\theta}\right)-\tilde{\varrho}\varrho r\sin\theta\left(\frac{f\left(\psi\right)}{u_{t}}-u^{\phi}L'\left(\psi\right)\right)+\left[\left(\partial_{r}\chi_{0}+\frac{\chi_{0}}{r}+\frac{c_{\theta}}{r}\right)\partial_{r}+\frac{1}{r^{2}}\left(\partial_{\theta}\chi_{0}+\cot\theta\chi_{0}-rc_{r}\right)\right]\partial_{\theta}\ln\tilde{\varrho}.\]
The two components of the $2$-vector source $S_{M}^{m}$ are the
$r$-component\begin{eqnarray*}
S_{M}^{r} & = & A^{2}N^{2}\left\{ 4\left(\alpha+\nu\right)_{,r}\left[M^{r}\left(\alpha+\nu\right)_{,r}+M_{\;\;,r}^{r}\right]+\frac{4}{r^{2}}\left(\alpha+\nu\right)_{,\theta}\left[M^{r}\left(\alpha+\nu\right)_{,\theta}+M_{\;\;,\theta}^{r}\right]\right.\\
 &  & \;\;\;\;\;\;\;\;\;\;\;\left.-\frac{4}{r}M^{\theta}\left(\alpha+\nu\right)_{,\theta}+{S'}_{M}^{r}+2M^{r}S_{\alpha}\right\} \end{eqnarray*}
and the $\theta$-component\begin{eqnarray*}
S_{M}^{\theta} & = & A^{2}N^{2}\left\{ 4\left(\alpha+\nu\right)_{,r}\left[M^{\theta}\left(\alpha+\nu\right)_{,r}+M_{\;\;,r}^{\theta}\right]+\frac{4}{r^{2}}\left(\alpha+\nu\right)_{,\theta}\left[M^{\theta}\left(\alpha+\nu\right)_{,\theta}+M_{\;\;,\theta}^{\theta}\right]\right.\\
 &  & \;\;\;\;\;\;\;\;\;\;\;\left.+\frac{4}{r}M^{\theta}\left(\alpha+\nu\right)_{,r}+\frac{4}{r^{3}}M^{r}\left(\alpha+\nu\right)_{,\theta}+{S'}_{M}^{\theta}+2M^{\theta}S_{\alpha}\right\} .\end{eqnarray*}
The three components of the $3$-vector source $S_{N}^{a}$ are the
$r$-component\begin{eqnarray*}
S_{N}^{r} & = & A^{2}\left\{ -16\pi NJ^{r}-2K^{rr}N_{,r}-2K^{r\theta}N_{,\theta}-2m^{r}m^{\theta}N_{\;\;,r\theta}^{r}-N_{\;,r}^{r}\left\{ 2\left[\frac{1}{A^{2}}+\left(m^{r}\right)^{2}\right]\alpha_{,r}+m^{r}m^{\theta}\left(2\alpha_{,\theta}-\mu_{,\theta}\right)\right.\right.\\
 &  & \;\;\;\;\;\;\;+\left.m^{r}\frac{M_{\;,\theta}^{\theta}}{M}+m^{\theta}\frac{M_{\;\;,\theta}^{r}}{M}\right\} -N_{\;,\theta}^{r}\left\{ 2\left[\frac{1}{\left(rA\right)^{2}}+2\left(m^{\theta}\right)^{2}\right]\alpha_{,\theta}+m^{r}m^{\theta}\left(\mu_{,r}+4\alpha_{,r}+\frac{1}{r}\right)-\frac{m^{r}M_{\;,\theta}^{r}}{r^{2}M}\right.\\
 &  & \;\;\;\;\;\;\;+\left.\frac{m^{\theta}}{M}\left(M_{\;\;,r}^{r}+2M_{\;\;,\theta}^{\theta}\right)\right\} -2N_{\;,r}^{\theta}\left[\frac{\alpha_{,\theta}}{A^{2}}+\left(m^{r}\right)^{2}\mu_{,\theta}-m^{r}\frac{M_{\;\;,\theta}^{r}}{M}\right]-N_{\;,\theta}^{\theta}\left\{ -2\left[\frac{1}{A^{2}}+\left(m^{r}\right)^{2}\right]\alpha_{,r}\right.\\
 &  & \;\;\;\;\;\;\;+\left.2m^{r}m^{\theta}\left(\mu_{,\theta}-\alpha_{,\theta}\right)-2m^{r}\frac{M_{\;,\theta}^{\theta}}{M}+\frac{m^{\theta}}{M}\left(r^{2}M_{\;\;,r}^{\theta}-M_{\;\;,\theta}^{r}\right)\right\} -N_{\;\;,\theta}^{\varphi}\left\{ \frac{2M^{r}}{\left(rA\right)^{2}}\left(\mu_{,\theta}-\alpha_{,\theta}\right)+2\frac{M^{\theta}}{A^{2}}\left[\alpha_{,r}-\mu_{,r}\right.\right.\\
 &  & \;\;\;\;\;\;\;+\left.\frac{1}{r}+\frac{\left(Am^{r}\right)^{2}}{r}\right]+2m^{r}m^{\theta}\left(M_{\;\;,\theta}^{\theta}-M_{\;\;,r}^{r}\right)-\frac{1}{r^{2}}\left[\frac{1}{A^{2}}-\left(m^{r}\right)^{2}+\left(rm^{\theta}\right)^{2}\right]M_{\;\;,\theta}^{r}+\left[\frac{1}{A^{2}}+\left(m^{r}\right)^{2}\right.\\
 &  & \;\;\;\;\;\;\;-\left.\left.\left(rm^{\theta}\right)^{2}\right]M_{\;\;,r}^{\theta}\right\} -N^{r}\left\{ 2\left[\frac{1}{A^{2}}+2\left(m^{r}\right)^{2}\right]\alpha_{,r}\left(\mu_{,r}-\frac{1}{r}\right)-\left[\frac{1}{A^{2}}-\left(m^{r}\right)^{2}\right]\frac{M_{,rr}}{M}+m^{r}m^{\theta}\left(-\mu_{,r}\mu_{,\theta}\right.\right.\\
 &  & \;\;\;\;\;\;\;+\left.4\alpha_{,\theta}\mu_{,r}-2\frac{\alpha_{,\theta}}{r}-2\alpha_{,r}\alpha_{,\theta}+\frac{M_{,r\theta}}{M}-2\frac{A_{,r\theta}}{A}\right)+2\left(m^{r}\right)^{2}\left[\frac{\mu_{,r}}{r}-\left(\alpha_{,r}\right)^{2}-\frac{A_{,rr}}{A}\right]+\frac{m^{r}}{M}\left[2M_{\;\;,r}^{r}\left(\mu_{,r}-4\alpha_{,r}\right.\right.\\
 &  & \;\;\;\;\;\;\;-\left.\left.\frac{1}{r}\right)-M_{\;\;,r}^{\theta}\left(\mu_{,\theta}+2\alpha_{,\theta}\right)+2M_{\;\;,\theta}^{\theta}\left(\mu_{,r}-\alpha_{,r}-\frac{1}{r}\right)-2M_{\;\;,rr}^{r}-M_{\;\;,r\theta}^{\theta}\right]+\frac{m^{\theta}}{M}\left[M_{\;\;,r}^{r}\left(\mu_{,\theta}-4\alpha_{,\theta}\right)\right.\\
 &  & \;\;\;\;\;\;\;-\left.\left.M_{\;\;,r\theta}^{r}\right]-\frac{1}{M^{2}}\left[2\left(M_{\;\;,r}^{r}\right)^{2}+M_{\;\;,r}^{r}M_{\;\;,\theta}^{\theta}+r^{2}\left(M_{\;\;,r}^{\theta}\right)^{2}\right]\right\} -N^{\theta}\left\{ \frac{2}{A^{2}}\alpha_{,\theta}\mu_{,r}+m^{r}m^{\theta}\left[4\alpha_{,\theta}\mu_{,\theta}-2\left(\alpha_{,\theta}\right)^{2}\right.\right.\\
 &  & \;\;\;\;\;\;\;-\left.\left(\mu_{,\theta}\right)^{2}+\frac{M_{,\theta\theta}}{M}-2\frac{A_{,\theta\theta}}{A}\right]+2\left(m^{r}\right)^{2}\left(\frac{\mu_{,\theta}}{r}-\frac{\alpha_{,\theta}}{r}-\alpha_{,r}\alpha_{,\theta}+2\mu_{,\theta}\alpha_{,r}-\frac{A_{,r\theta}}{A}\right)+\frac{m^{r}}{M}\left[2M_{\;\;,r}^{r}\left(\mu_{,\theta}-\alpha_{,\theta}\right)\right.\\
 &  & \;\;\;\;\;\;\;-\left.2M_{\;\;,\theta}^{r}\left(3\alpha_{,r}+\frac{1}{r}\right)+M_{\;\;,\theta}^{\theta}\left(\mu_{,\theta}-4\alpha_{,\theta}\right)-M_{\;\;,\theta\theta}^{\theta}-2M_{\;\;,r\theta}^{r}\right]+\frac{m^{\theta}}{M}\left[M_{\;\;,\theta}^{r}\left(\mu_{,\theta}-4\alpha_{,\theta}\right)-M_{\;\;,\theta\theta}^{r}\right]\\
 &  & \;\;\;\;\;\;\;-\left.\frac{1}{M^{2}}\left(M_{\;\;,\theta}^{r}M_{\;\;,\theta}^{\theta}+2M_{\;\;,r}^{r}M_{\;\;,\theta}^{r}+r^{2}M_{\;\;,r}^{\theta}M_{\;\;,\theta}^{\theta}\right)\right\} -\left(m^{r}\right)^{2}\left\{ N_{\;\;,rr}^{r}+\left(\frac{1}{r}+\mu_{,r}\right)N_{\;\;,r}^{r}-\left[\frac{1}{r^{2}}+\left(\mu_{,r}\right)^{2}\right]N^{r}\right\} \\
 &  & \;\;\;\;\;\;\;-\left.\left(m^{\theta}\right)^{2}N_{\;\;,\theta\theta}^{r}+\left(m^{\theta}\right)^{2}\mu_{,\theta}N_{\;\;,\theta}^{r}+\frac{2}{r}\left(m^{r}\right)^{2}N_{\;\;,\theta}^{\theta}-\left(m^{r}\right)^{2}\beta_{,r\theta}N^{\theta}\right\} \\
 &  & -\beta_{,r}N_{\;\;,r}^{r}+\left[\frac{2}{r}\beta_{,r}+\left(\beta_{,r}\right)^{2}\right]N^{r}-\frac{1}{r^{2}}\beta_{,\theta}N_{\;\;,\theta}^{r}+\left(\frac{2}{r}\beta_{,\theta}+2\cot\theta\beta_{,r}+\beta_{,r\theta}+2\beta_{,r}\beta_{,\theta}\right)N^{\theta},\end{eqnarray*}
the $\theta$-component\begin{eqnarray*}
S_{N}^{\theta} & = & A^{2}\left\{ -16\pi NJ^{\theta}-2K^{\theta r}N_{,r}-2K^{\theta\theta}N_{,\theta}-2m^{r}m^{\theta}N_{\;\;,r\theta}^{\theta}-N_{\;\;,r}^{r}\left\{ -2\left[\frac{1}{\left(rA\right)^{2}}+\left(m^{\theta}\right)^{2}\right]\alpha_{,\theta}+2m^{r}m^{\theta}\left(\mu_{,r}-\alpha_{,r}\right)\right.\right.\\
 &  & \;\;\;\;\;\;\;+\left.\frac{1}{M}\left[m^{r}\left(\frac{M_{\;\;,\theta}^{r}}{r^{2}}-M_{\;\;,r}^{\theta}\right)-2m^{\theta}M_{\;\;,r}^{r}\right]\right\} -N_{\;\;,\theta}^{r}\left[\frac{2}{\left(rA\right)^{2}}\alpha_{,r}+2\left(m^{\theta}\right)^{2}\mu_{,r}-2m^{\theta}\frac{M_{\;\;,r}^{\theta}}{M}\right]-N_{\;\;,r}^{\theta}\left\{ 2\left[\frac{1}{A^{2}}\right.\right.\\
 &  & \;\;\;\;\;\;\;+\left.\left.2\left(m^{r}\right)^{2}\right]\alpha_{,r}+m^{r}m^{\theta}\left(\mu_{,\theta}+4\alpha_{,\theta}\right)+\frac{1}{M}\left[m^{r}\left(2M_{\;\;,r}^{r}+M_{\;\;,\theta}^{\theta}\right)-m^{\theta}r^{2}M_{\;\;,r}^{\theta}\right]\right\} -N_{\;\;,\theta}^{\theta}\left\{ 2\left[\frac{1}{\left(rA\right)^{2}}\right.\right.\\
 &  & \;\;\;\;\;\;\;+\left.\left.\left(m^{\theta}\right)^{2}\right]\alpha_{,\theta}+m^{r}m^{\theta}\left(2\alpha_{,r}-\mu_{,r}+\frac{1}{r}\right)+\frac{1}{M}\left(m^{r}M_{\;\;,r}^{\theta}+m^{\theta}M_{\;\;,r}^{r}\right)\right\} -N_{\;\;,r}^{\varphi}\left\{ \frac{2M^{r}}{\left(rA\right)^{2}}\left(\alpha_{,\theta}-\mu_{,\theta}\right)\right.\\
 &  & \;\;\;\;\;\;\;+2\frac{M^{\theta}}{A^{2}}\left[\mu_{,r}-\alpha_{,r}-\frac{1}{r}-\frac{\left(Am^{r}\right)^{2}}{r}\right]+2m^{r}m^{\theta}\left(M_{\;\;,r}^{r}-M_{\;\;,\theta}^{\theta}\right)+\frac{1}{r^{2}}\left[\frac{1}{A^{2}}-\left(m^{r}\right)^{2}+\left(rm^{\theta}\right)^{2}\right]M_{\;\;,\theta}^{r}-\left[\frac{1}{A^{2}}\right.\\
 &  & \;\;\;\;\;\;\;+\left.\left.\left(m^{r}\right)^{2}-\left(rm^{\theta}\right)^{2}\right]M_{\;\;,r}^{\theta}\right\} -N^{r}\left\{ -\left(m^{\theta}\right)^{2}\mu_{,r}\mu_{,\theta}+\left(m^{\theta}\right)^{2}\frac{M_{,r\theta}}{M}-\frac{1}{\left(rA\right)^{2}}\left(2\cot\theta\beta_{,r}+\beta_{,r\theta}+2\beta_{,r}\beta_{,\theta}\right)\right.\\
 &  & \;\;\;\;\;\;\;+\frac{2}{r}\left[\frac{1}{\left(rA\right)^{2}}-\left(m^{\theta}\right)^{2}\right]\alpha_{,\theta}+\frac{2}{\left(rA\right)^{2}}\alpha_{,r}\mu_{,\theta}+m^{r}m^{\theta}\left[2\frac{\mu_{,r}}{r}-4\frac{\alpha_{,r}}{r}-\frac{1}{r^{2}}+4\alpha_{,r}\mu_{,r}-2\left(\alpha_{,r}\right)^{2}-\left(\mu_{,r}\right)^{2}+\frac{M_{,rr}}{M}\right.\\
 &  & \;\;\;\;\;\;\;-\left.2\frac{A_{,rr}}{A}\right]+2\left(m^{\theta}\right)^{2}\left[\left(2\mu_{,r}-\alpha_{,r}\right)\alpha_{,\theta}-\frac{A_{,r\theta}}{A}\right]+\frac{m^{r}}{M}\left[\left(\mu_{,r}-4\alpha_{,r}-\frac{3}{r}\right)M_{\;\;,r}^{\theta}-M_{\;\;,rr}^{\theta}\right]+\frac{m^{\theta}}{M}\left[M_{\;\;,r}^{r}\left(\mu_{,r}\right.\right.\\
 &  & \;\;\;\;\;\;\;-\left.\left.4\alpha_{,r}-\frac{1}{r}\right)-6M_{\;\;,r}^{\theta}\alpha_{,\theta}-2M_{\;\;,\theta}^{\theta}\left(-\mu_{,r}+\alpha_{,r}+\frac{1}{r}\right)-M_{\;\;,rr}^{r}-2M_{\;\;,r\theta}^{\theta}\right]-\frac{1}{M^{2}}\left(\frac{1}{r^{2}}M_{\;\;,r}^{r}M_{\;\;,\theta}^{r}\right.\\
 &  & \;\;\;\;\;\;\;+\left.\left.M_{\;\;,r}^{r}M_{\;\;,r}^{\theta}+2M_{\;\;,r}^{\theta}M_{\;\;,\theta}^{\theta}\right)\right\} -N^{\theta}\left\{ 2\left[\frac{1}{\left(rA\right)^{2}}+2\left(m^{\theta}\right)^{2}\right]\alpha_{,\theta}\mu_{,\theta}+m^{r}m^{\theta}\left[2\frac{\mu_{,\theta}}{r}-2\frac{\alpha_{,\theta}}{r}-\mu_{,r}\mu_{,\theta}+4\alpha_{,r}\mu_{,\theta}\right.\right.\\
 &  & \;\;\;\;\;\;\;-\left.2\alpha_{,r}\alpha_{,\theta}+\frac{M_{,r\theta}}{M}-2\frac{A_{,r\theta}}{A}\right]-2\left(m^{\theta}\right)^{2}\left[\left(\alpha_{,\theta}\right)^{2}+\frac{A_{,\theta\theta}}{A}\right]+\frac{m^{r}}{M}\left[M_{\;\;,\theta}^{\theta}\left(\mu_{,r}-4\alpha_{,r}-\frac{3}{r}\right)-M_{\;\;,r\theta}^{\theta}\right]\\
 &  & \;\;\;\;\;\;\;+\frac{m^{\theta}}{M}\left[2M_{\;\;,r}^{r}\left(\mu_{,\theta}-\alpha_{,\theta}\right)-M_{\;\;,\theta}^{r}\left(\mu_{,r}+2\alpha_{,r}+\frac{1}{r}\right)+2M_{\;\;,\theta}^{\theta}\left(\mu_{,\theta}-4\alpha_{,\theta}\right)-M_{\;\;,r\theta}^{r}-2M_{\;\;,\theta\theta}^{\theta}\right]\\
 &  & \;\;\;\;\;\;\;-\left.\frac{1}{M^{2}}\left[\left(\frac{M_{\;\;,\theta}^{r}}{r}\right)^{2}+2\left(M_{\;\;,\theta}^{\theta}\right)^{2}+M_{\;\;,r}^{r}M_{\;\;,\theta}^{\theta}\right]\right\} -\left(m^{r}\right)^{2}N_{\;\;,rr}^{\theta}-\frac{1}{A^{2}}\beta_{,r}N_{\;\;,r}^{\theta}-\left(m^{r}\right)^{2}\left(\frac{3}{r}-\mu_{,r}\right)N_{\;\;,r}^{\theta}\\
 &  & \;\;\;\;\;\;\;-\left.\left(m^{\theta}\right)^{2}\left(N_{\;\;,\theta\theta}^{\theta}+\mu_{,\theta}N_{\;\;,\theta}^{\theta}\right)+\left(m^{\theta}\right)^{2}\left(\cot^{2}\theta+1-\beta_{,\theta\theta}\right)N^{\theta}\right\} \\
 &  & -\frac{1}{r^{2}}\beta_{,\theta}N_{\;\;,\theta}^{\theta}+N^{\theta}\frac{1}{r^{2}}\left[4\cot\theta\beta_{,\theta}+2\left(\beta_{,\theta}\right)^{2}+\beta_{,\theta\theta}\right]\end{eqnarray*}
and the $\phi$-component\begin{eqnarray*}
S_{N}^{\phi} & = & A^{2}\left\{ -16\pi NJ^{\varphi}-2K^{\varphi r}N_{,r}-2K^{\varphi\theta}N_{,\theta}-2m^{r}m^{\theta}N_{\;\;,r\theta}^{\varphi}-2\frac{N_{\;\;,r}^{r}}{M}\left[m^{r}\left(\mu_{,r}-\alpha_{,r}\right)-m^{\theta}\alpha_{,\theta}-\frac{M_{\;\;,r}^{r}}{M}\right]\right.\\
 &  & \;\;\;\;\;\;\;-\frac{N_{\;\;,\theta}^{r}}{M}\left[2m^{\theta}\mu_{,r}-\frac{1}{M}\left(M_{\;\;,r}^{\theta}+\frac{M_{\;\;,\theta}^{r}}{r^{2}}\right)\right]-\frac{N_{\;\;,r}^{\theta}}{M}\left[2m^{r}\mu_{,\theta}-\frac{1}{M}\left(r^{2}M_{\;\;,r}^{\theta}+M_{\;\;,\theta}^{r}\right)\right]-2\frac{N_{\;\;,\theta}^{\theta}}{M}\left[m^{\theta}\left(\mu_{,\theta}-\alpha_{,\theta}\right)\right.\\
 &  & \;\;\;\;\;\;\;-\left.m^{r}\left(\alpha_{,r}+\frac{1}{r}\right)-\frac{M_{\;\;,\theta}^{\theta}}{M}\right]-N_{\;\;,r}^{\varphi}\left[-m^{r}m^{\theta}\mu_{,\theta}+4m^{r}\left(m^{r}\alpha_{,r}+m^{\theta}\alpha_{,\theta}\right)+\frac{m^{r}}{M}\left(4M_{\;\;,r}^{r}+M_{\;\;,\theta}^{\theta}\right)\right.\\
 &  & \;\;\;\;\;\;\;+\left.\frac{m^{\theta}}{M}\left(2M_{\;\;,\theta}^{r}+r^{2}M_{\;\;,r}^{\theta}\right)\right]-N_{\;\;,\theta}^{\varphi}\left[-m^{r}m^{\theta}\mu_{,r}+4m^{\theta}\left(m^{r}\alpha_{,r}+m^{\theta}\alpha_{,\theta}\right)+\frac{3}{r}m^{r}m^{\theta}+\frac{m^{r}}{M}\left(\frac{M_{\;\;,\theta}^{r}}{r^{2}}+2M_{\;\;,r}^{\theta}\right)\right.\\
 &  & \;\;\;\;\;\;\;+\left.\frac{m^{\theta}}{M}\left(M_{\;\;,r}^{r}+4M_{\;\;,\theta}^{\theta}\right)\right]-\frac{N^{r}}{M}\left\{ m^{r}\left[2\frac{\mu_{,r}}{r}-4\frac{\alpha_{,r}}{r}-\frac{1}{r^{2}}-\left(\mu_{,r}\right)^{2}-2\left(\alpha_{,r}\right)^{2}+4\alpha_{,r}\mu_{,r}+\frac{M_{,rr}}{M}-2\frac{A_{,rr}}{A}\right]\right.\\
 &  & \;\;\;\;\;\;\;+m^{\theta}\left(-\mu_{,r}\mu_{,\theta}+4\alpha_{,\theta}\mu_{,r}-2\frac{\alpha_{,\theta}}{r}-2\alpha_{,r}\alpha_{,\theta}+\frac{M_{,r\theta}}{M}-2\frac{A_{,r\theta}}{A}\right)+\frac{1}{M}\left[M_{\;\;,r}^{r}\left(\mu_{,r}-4\alpha_{,r}-\frac{1}{r}\right)-M_{\;\;,r}^{\theta}\left(\mu_{,\theta}\right.\right.\\
 &  & \;\;\;\;\;\;\;+\left.\left.\left.2\alpha_{,\theta}\right)+2M_{\;\;,\theta}^{\theta}\left(\mu_{,r}-\alpha_{,r}-\frac{1}{r}\right)-M_{\;\;,rr}^{r}-M_{\;\;,r\theta}^{\theta}\right]\right\} -\frac{N^{\theta}}{M}\left\{ m^{r}\left[-\mu_{,r}\mu_{,\theta}+\left(4\alpha_{,r}+\frac{2}{r}\right)\mu_{,\theta}-2\frac{\alpha_{,\theta}}{r}\right.\right.\\
 &  & \;\;\;\;\;\;\;-\left.2\alpha_{,r}\alpha_{,\theta}+\frac{M_{,r\theta}}{M}-2\frac{A_{,r\theta}}{A}\right]+m^{\theta}\left[-\left(\mu_{,\theta}\right)^{2}-2\left(\alpha_{,\theta}\right)^{2}+4\alpha_{,\theta}\mu_{,\theta}+\frac{M_{,\theta\theta}}{M}-2\frac{A_{,\theta\theta}}{A}\right]+\frac{1}{M}\left[2M_{\;\;,r}^{r}\left(\mu_{,\theta}-\alpha_{,\theta}\right)\right.\\
 &  & \;\;\;\;\;\;\;-\left.\left.M_{\;\;,\theta}^{r}\left(\mu_{,r}+2\alpha_{,r}+\frac{1}{r}\right)+M_{\;\;,\theta}^{\theta}\left(\mu_{,\theta}-4\alpha_{,\theta}\right)-M_{\;\;,r\theta}^{r}-M_{\;\;,\theta\theta}^{\theta}\right]\right\} -\left(m^{r}\right)^{2}N_{\;\;,rr}^{\varphi}-\left[\frac{3}{A^{2}}-\left(m^{r}\right)^{2}\right]\beta_{,r}N_{\;\;,r}^{\varphi}\\
 &  & \;\;\;\;\;\;\;-\left.\left(m^{\theta}\right)^{2}N_{\;\;,\theta\theta}^{\varphi}-\left[\frac{3}{\left(rA\right)^{2}}-\left(m^{\theta}\right)^{2}\right]\beta_{,\theta}N_{\;\;,\theta}^{\varphi}+\left(m^{\theta}\right)^{2}\cot\theta N_{\;\;,\theta}^{\varphi}\right\} .\end{eqnarray*}
\end{widetext}

\bibliographystyle{apsrev4-1}
\bibliography{GRNS}

\end{document}